\documentclass[conference]{IEEEtran}
\IEEEoverridecommandlockouts
\usepackage{optidef}
\usepackage{authblk}
\usepackage{cite}
\usepackage{booktabs}
\usepackage{tabu}
\usepackage{amsmath,amssymb,amsfonts}
\usepackage[hang,flushmargin]{footmisc}
\usepackage{multirow}
\usepackage{float}
\usepackage{textcomp}
\usepackage{lscape}
\usepackage{url}
\usepackage{enumerate}
\usepackage{caption}
\usepackage{subcaption}
\usepackage{algpseudocode}
\usepackage{graphicx}
\usepackage{textcomp}
\usepackage{xcolor}
\usepackage{array}
\usepackage{mathtools,algorithm}

\usepackage{enumitem}

\newcommand{\CASE}[1]{\STATE \textbf{case} #1\textbf{:} \begin{ALC@g}}
\newcommand{\ENDCASE}{\end{ALC@g}}

\newcommand{\DEFAULT}{\STATE \textbf{default:} \begin{ALC@g}}
\newcommand{\ENDDEFAULT}{\end{ALC@g}}
\newcommand{\DEFAULTLINE}[1]{\STATE \textbf{default:} }
\newcolumntype{L}[1]{>{\raggedright\let\newline\\\arraybackslash\hspace{0pt}}m{#1}}
\newcolumntype{C}[1]{>{\centering\let\newline\\\arraybackslash\hspace{0pt}}m{#1}}
\newcolumntype{R}[1]{>{\raggedleft\let\newline\\\arraybackslash\hspace{0pt}}m{#1}}
\def\BibTeX{{\rm B\kern-.05em{\sc i\kern-.025em b}\kern-.08em
    T\kern-.1667em\lower.7ex\hbox{E}\kern-.125emX}}

\begin{document}

\title{Availability-Aware Dynamic RSA with Protection using Consecutive Sub-Channels}

\author[ ]{Varsha Lohani}
\author[ ]{Anjali Sharma}
\author[ ]{Yatindra Nath Singh} 

\affil[  ]{Department of Electrical Engineering, Indian Institute of Technology Kanpur, Kanpur, India}

\affil[  ]{\textit {lohani.varsha7@gmail.com*, anjalienix05@gmail.com and ynsingh@iitk.ac.in}}

\maketitle
\begin{abstract}
Flexible grid Optical Networks provide efficient spectrum utilization by employing the mechanisms to provide flexibility in the optical channel (spectrum slot) sizes. One of the research problems in Flexible grid Optical Networks is its survivability against failure. On the other hand, p-Cycles have not found practical use due to significant compute time required for finding optimal configuration for the size of networks seen in real-life. Therefore, for real-time scenarios, we can write heuristics which can assign protection to the new working paths without disturbing the existing traffic on all the other routes in flexible grid networks. The provisioning of protection to each link or path of the lightpath requests can be done using Dynamic Cycles (D-Cycles) or Dynamic Shared Backup Resource Protection (D-SBRP). However, protecting each link or path can lead to the wastage of the resources in the network. 
\end{abstract}

\section{Introduction}

p-Cycles offer ring-like switching speed and mesh-like spare capacity efficiency for providing protection against link failures. They can also protect all the links in a network against simultaneous failures of multiple links \cite{2L}. But it has been mostly studied for single link failure scenarios in the networks with the objective to minimize spare capacity under the condition of $100\%$ protection for any single link failure scenario. 

p-Cycles have not found practical use, only due to significant compute time required for finding optimal configurations for the size of networks seen in real-life. Therefore, \cite{sg} investigated the sub-graphing method to reduce the compute time for finding the p-cycles. Partitioning of network was used for finding the real-time protection provisioning. It was shown that protection can be achieved much more quickly as a consequence of the sub-graphing. 

Another alternative approach to achieve real-time protection is to make local information-based decisions using heuristics at all the nodes. The algorithm from \cite{sg} can be compared with such actual real-time scenarios where the protection is computed heuristically on need basis. In dynamically maintained protection, when a p-cycle is not protecting any working path, it can be dismantle to release the capacity. The existing p-cycles should be used whenever possible with suitable modifications. In the considered scenario, it is possible that we can set up a working path but cannot provide protection to all of its constituent links. This constraint exists because of limited spare capacity in practical networks, though the p-cycles in the network can be merged or expanded to minimize the required protection capacity. 

We expect that the proposed scheme may not always work with optimum capacity but can be pushed towards optimality as the network operates. While by partitioning, we can only periodically compute the optimal configuration and reconfigure the operating p-cycles to achieve optimality. The dynamic creation, dismantling, merger, and expansion of p-cycles are not done while using partitioning method, thus achieving a sub-optimal performance most of the time except at the instants just after the periodic re-configurations. Also, the ILP for a partition should consider the total fixed capacity in all the links for both working and protection capacities. The objective of ILP will change to provide maximum protection within the available resources, but sometimes the optimal solution may not provide $100 \%$ single fault tolerance. This is the problem one will face while using ILP based solution in real-time network operation.

Alternatively, for real-time scenarios, we can write heuristics where we can assign protection to the new working paths without disturbing the existing paths and the traffic carried by them. The provisioning of protection to each of the links or the whole path assigned to a lightpath requests can be done using Dynamic Cycles (D-Cycles) or Dynamic Shared Backup Resource Protection (D-SBRP). However, protecting each link or path can lead to the wastage of the resources in the network.

An alternative approach for a real-time situation is to provide protection based on the \textbf{Availability} of the link or path.  Availability is the probability that the Path or links are working at any random time \textit{t}. Availability decides the resilience of the networks. In other words, if the Availability of a link or path is higher than some threshold value, then there is no need for protection. However, if the Availability is lower than the threshold, then protecting some of the links can improve the overall Availability of the path. This approach provides protection, while requiring lesser spare capacity\footnote{it means less contribution from the total capacity to the spare capacity.}. However, there is no guarantee of  $100 \% $ restorability.  

\section{Problem Statement} 

Protection means apriori assignment of spare capacity, to be used in case of link or node failure in the middle of a call.  Backup paths setup using spare capacity provide protection to restore the communication in the event of failure. p-Cycles have been considered for protection against link failure in quite a good amount of literature. They are one of the best methods to provide most efficient protection and achieve much faster restoration speeds in case of failures. However, computing p-cycle apriori normally requires large computation time. In addition, if sufficient resources are not there to setup the p-cycles, the call cannot be protected, and there is wastage of time in attempting to compute them. A similar problem occurs with Failure-Independent Path Protecting (FIPP) p-cycles, which provision pre-computed and pre-connected protection to the working paths \cite{fippa}. Therefore, we propose a method for protection in which cycles are provisioned on the fly dynamically in the networks on need basis. 

Another parameter of interest considered in this paper is the Availability of optical links and hence the path formed with the links, before provisioning for the protection. Chen \textit{et. al} \cite{ava1} proposed analysis for service availability using p-cycles protection techniques. They proposed a theoretical model on the service availability when using p-cycles, considering all the possible scenarios. The algorithms were based on the already proposed ones, and were compared with the existing algorithms for p-cycle design. Algorithms for availability analysis using Dedicated Path Protection and Shared Backup Path Protection \cite{ava2} were also proposed. They analyzed the service availability of various protection systems in EONs theoretically and then suggested an algorithm that could adjust the path protection scheme to meet various service availability needs. In \cite{ava3}, mathematical modeling of the Availability-aware FIPP p-cycles in Flexi- grid optical networks was given and cyclic partitioning was proposed for protection provisioning. 

Some of the research works \cite{avasfc} have used availability analysis for validation of their work related to protection provisioning. In most of the works, protection arrangements (and hence capacity) are computed even before a lightpath request arrives. The protection paths are mostly computed considering distance or hops as a parameter for computation. 

\section{Proposed Solution}

In this paper, we consider consecutive spectrum slots (used for RSA in \cite{DRSA}) for backup path computation. For networks with dynamically arriving and departing traffic, we propose a method for link protection that is a modified form of p-Cycles and we call it Dynamic Cycles (D-cycles) as they are formed and dismantled dynamically on need basis. Similarly, for path protection, we are using Dynamic Shared Backup Path Protection with Spectrum Slots Sharing. However, provisioning protection to each link or path of the lightpath requests leads to wastage of spare capacity. Also, it increases blocking of the requests as load increases. Instead we have considered availability as a parameter for best-path selection and provisioning of protection. 

In this paper, first, we will discuss protection techniques used. Reliability and Availability of a link and path will be discussed in the subsequent section. The Availability of paths in various scenarios is explained with the help of examples. Finally, we will present the proposed algorithms and evaluate its effectiveness.

\section{Protection Techniques}

Optical networks need to carry a large volume of data traffic while maintaining service continuity even in the presence of faults. If not automatically protected and restored within a short time after failure, even a single link failure will result in loss of significant amount of data. Therefore, survivability against link or path failures is a necessary design requirement for high-speed optical networks. 

This paper first explores path protection for real-time scenarios, i.e., Dynamic Shared Backup Path and Spectrum Slots (DSBPSS) \cite{sbrp}. Total traffic from all the primary paths passing through a failed link is switched to backup protection paths whenever restoration is done.

Another technique for protection, proposed by us is \textbf{Dynamic Cycles (D-cycles)} for link protection. In link protection, the traffic through the faulty link is restored via another path joining the two endpoints of the link. The switching speed of link protection is faster than the path protection. Also, fault detection is localized in link protection. Further all paths passing through the failed links are restored as single event, unlike path protection. However, each protection switch over will require spectrum converters. 

\subsection{Dynamic Shared Backup Path and Spectrum Slots}
In Dynamic Shared Backup Path and Spectrum Slots (DSBPSS), a single backup path protects a number of disjoint primary working paths with an extra feature of sharing the spectrum slots. When a failure occurs in any one of the primary paths, then the failed path's traffic is switched to the backup path. The primary paths must be link-disjoint with the backup path as well as among themselves. After the setup lightpath are released, the backup paths are also dismantled.

\begin{figure}[h]
 \centering
    \includegraphics[width=\linewidth]{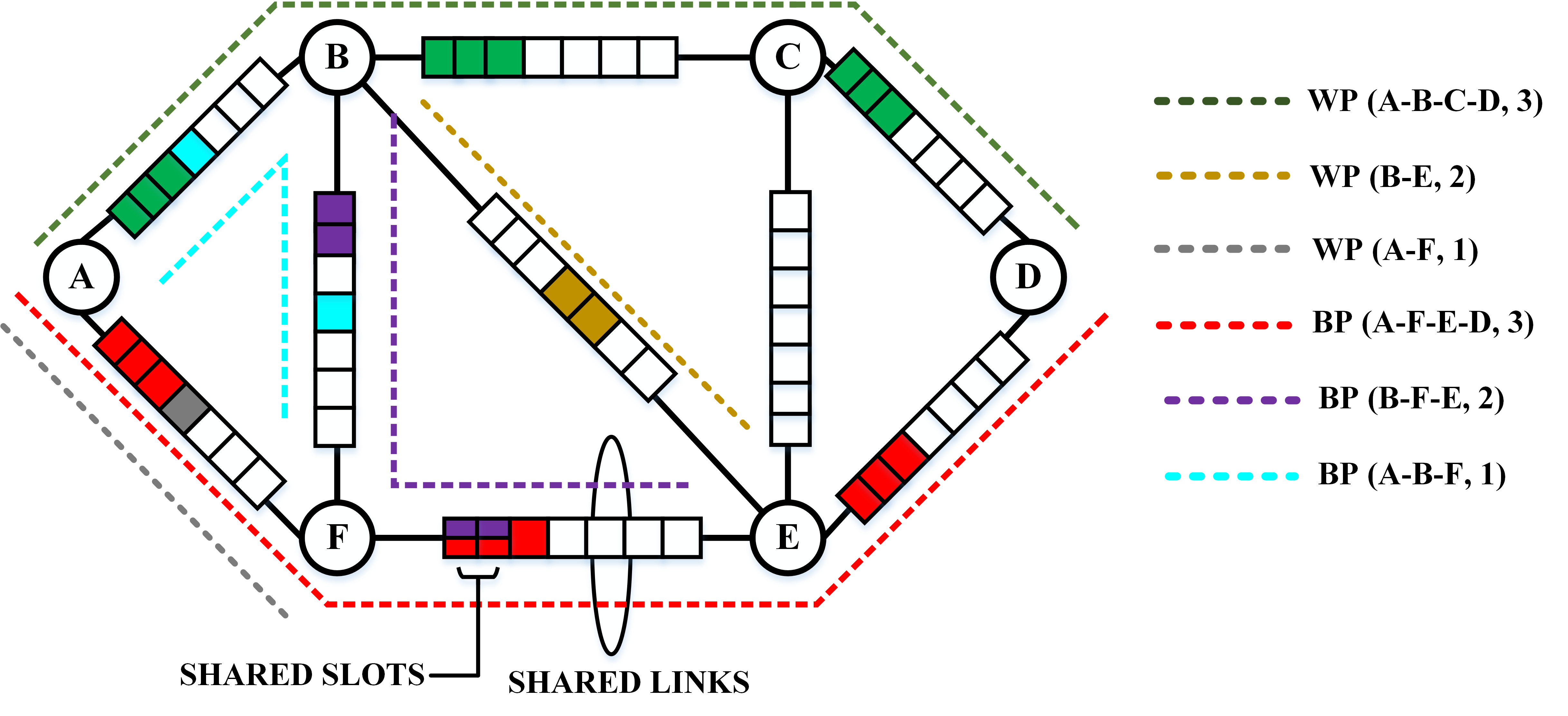}
    \caption{DSBPSS for three working paths.}
    \label{fig:sbppss}
\end{figure}

Figure \ref{fig:sbppss} shows an example of DSBPSS. There are three working lightpath requests A-B-C-D, B-E and A-F. For lightpath requests A-B-C-D with three slots, backup path A-F-E-D with three slots is provisioned. Whereas for lightpath request B-E with two slots, backup path B-F-E with two slots is provisioned, and for lightpath request A-F with a single slot, backup path A-B-F with one slot is provisioned. The link F-E is common for protection of two working paths such that the protection slots are shared by both the working paths. The links of the working paths are disjoint. 

Consequently, the failure of one of the working paths cannot affect the other. We consider the probability of the simultaneous failure of the two paths to be very low. Here, WP represents working path and BP is for Backup Path. Each of the paths is provisioned with the spectrum slots on disjoint backup paths. If the spectrum slots are not available on the backup path, another backup path is checked. If available, then it will be provisioned. Else the path remains unprotected.

\begin{itemize}
\item [\textbf{Pros}]
\item Backup Path can protect against the failure of both the intermediate nodes as well as the constituent link failures. The switchover happens at the source and destination nodes.   
\item There is no need for spectrum converters for switching from Working Path to Backup Path.
\item [\textbf{Cons}]
\item The switching speed of the path protection scheme is slower than the link protection schemes.
\item If there is a single link failure, the spectrum slots used on other links of the working paths are no more in use after the restoration. In other words, useful slots get wasted during restored operation. Though these slots are now available for other request setups.
\item When a link fails, all the paths passing through it are to be restored using backup paths. This creates a kind of sudden burst of multiple path configurations all across the network.
\end{itemize}

\subsubsection{Conditions for Backup Slots Assignment}
For backup slots assignment for protection while using RSA using DSBPSS, various conditions are required to be followed.  

\begin{figure}[h]
\begin{subfigure}{0.5\textwidth}
    \centering
    \includegraphics[width=\linewidth]{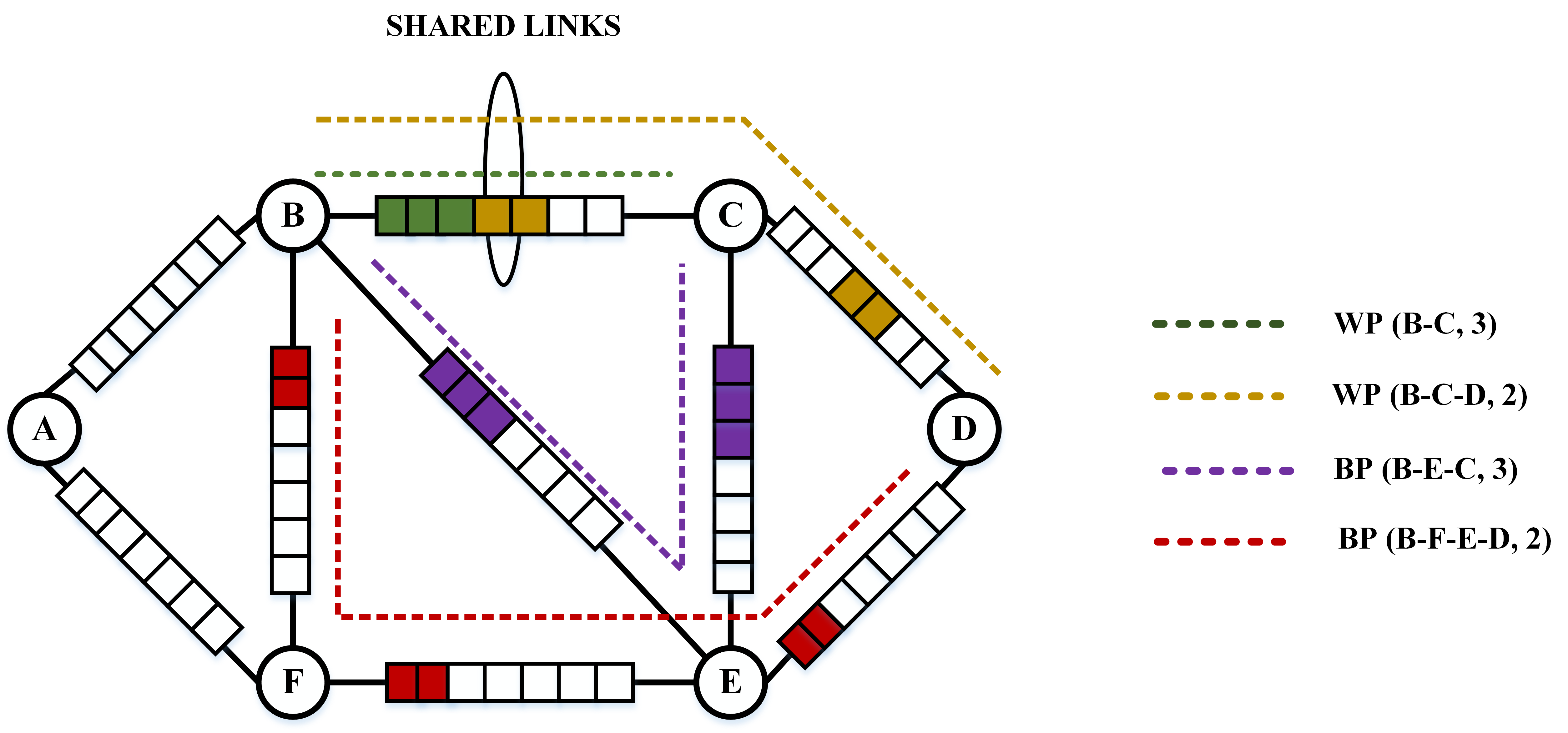}
    \caption{}
    \label{fig:c1}
\end{subfigure}
\hfill
\begin{subfigure}{0.5\textwidth}
    \centering
    \includegraphics[width=\linewidth]{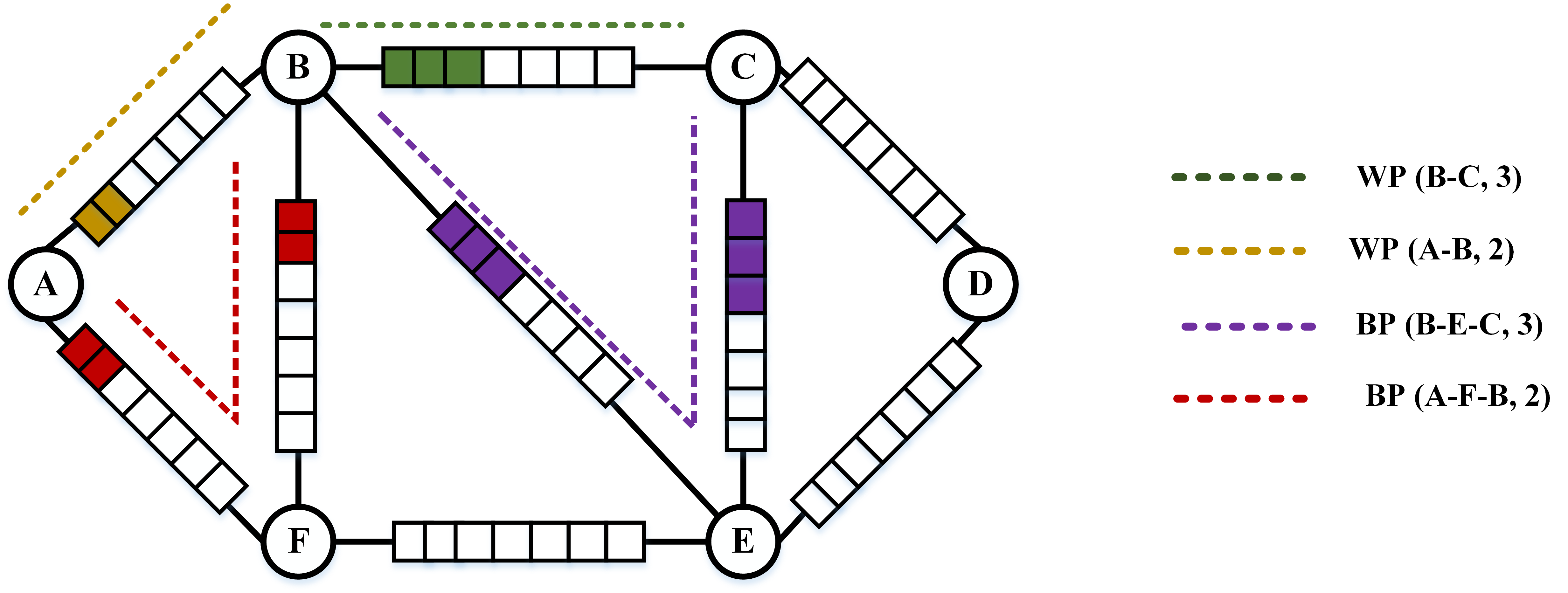}
    \caption{}
    \label{fig:c2}
\end{subfigure}
\hfill
\begin{subfigure}{0.5\textwidth}
    \centering
    \includegraphics[width=\linewidth]{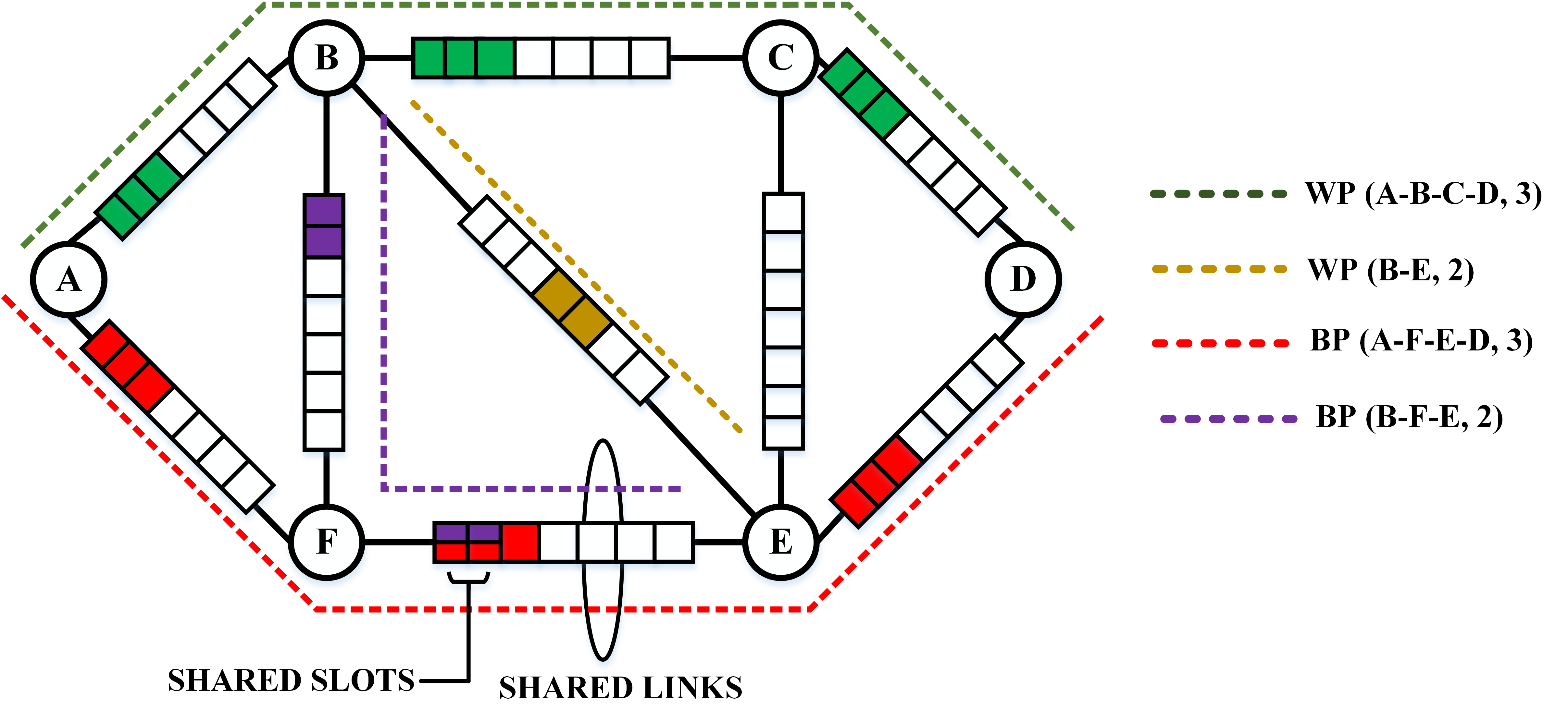}
    \caption{}
    \label{fig:c3}
\end{subfigure}
\hfill
\begin{subfigure}{0.5\textwidth}
    \centering
    \includegraphics[width=\linewidth]{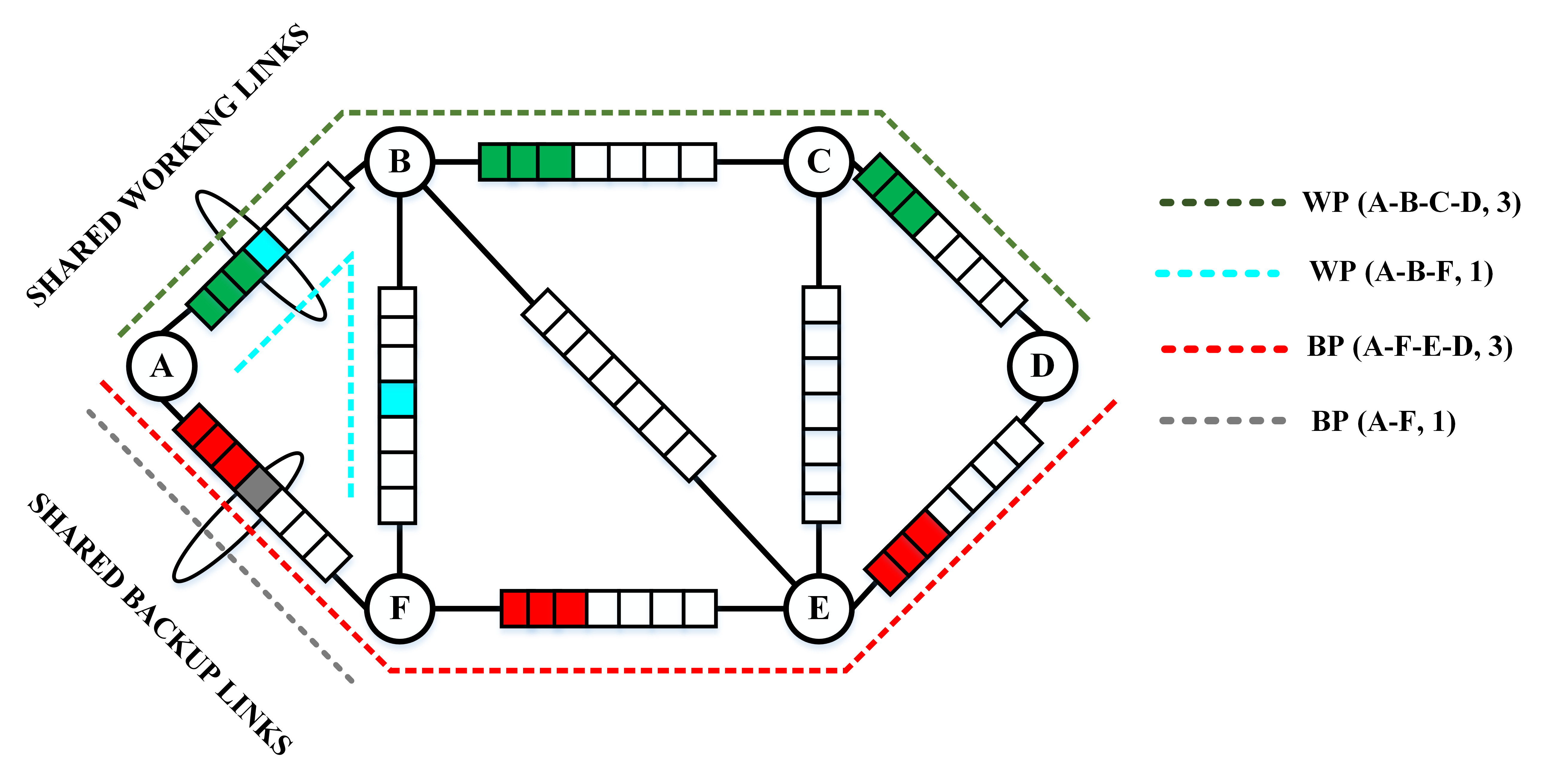}
    \caption{}
    \label{fig:c4}
\end{subfigure}
\caption{Backup Slot Assignment.}
\label{fig:cond}
\end{figure}

\subsubsection*{Condition I: Working paths have non-disjoint links.} Active working and Current working path requests share links and thus allotted different slots for working capacity, and their corresponding backup paths are link-disjoint among themselves as well as with working paths, thus the allocated backup slots are independent being on different backup paths as shown in figure \ref{fig:c1}. 

Let us consider active lightpath request, B-C with three slots requirement, and its corresponding backup path B-E-C is used for protection as shown in figure \ref{fig:c1}. Whereas current lightpath request B-C-D with two slots requirement, its corresponding backup path B-F-E-D can be used for protection. The link B-C of active and current lightpath request are shared. Therefore, the allocated backup links and slots must be disjoint otherwise only one of path or link is protected in the event of failure on link B-C. Here, WP represents working path and BP is for Backup Path.

\subsubsection*{Condition II: Working paths and Backup Paths are link-disjoint} Active working and Current working path requests are link-disjoint and their corresponding backup paths are also link-disjoint, then allocated backup slots are always disjoint being on different links. 

Let us consider active lightpath request, B-C with three slots requirement, and its corresponding backup path B-E-C is used for protection  as shown in figure \ref{fig:c2}. Whereas current lightpath request A-B with two slots requirement, its corresponding backup path A-F-B is used for protection. Since WPs and BPs are link disjoint therefore the allocated backup slots are also link-disjoint. 

\subsubsection*{Condition III: Working paths are link-disjoint and Backup paths links are non-disjoint} Active working and Current working path requests are link-disjoint and their corresponding backup paths share links, then allocated backup slots can be shared on the shared links.

Let us consider an active lightpath request, A-B-C-D with three slots requirement, and its corresponding backup path A-F-E-D is used for protection  as shown in figure \ref{fig:c3}. Whereas current lightpath request B-E is with two slots requirement, its corresponding backup path for protection is B-F-E. Since the WPs are link-disjoint, therefore, their BPs can share links as well as slots. 

\subsubsection*{Condition IV:  Working paths and Backup paths links are non-disjoint} Active working and Current working path requests share some links and assigned different slots on shared links. Their corresponding backup paths also share links, then slot shareability is not feasible on backup paths and disjoint slots should be considered for backup provisioning. 

Let us consider active lightpath request, A-B-C-D with three slot requirement, and its corresponding backup path A-F-E-D is used for protection as shown in figure \ref{fig:c4}. Whereas current lightpath request A-B-F is with one slots requirement; its corresponding backup path A-F is used for protection. Since the WPs and BPs are not link-disjoint. Therefore, the slots assigned for protection should be disjoint.

\subsection{Dynamic Cycles (D-Cycles)}

In \cite{DRSA}, we considered contiguous and continuous spectrum slots for Routing and Spectrum Assignment. A similar algorithm is also used in this paper for finding backup paths. Since we are using parameters that change with the network conditions, instead of pre-configured-Cycles, Dynamic cycles are used.

The D-cycles technique is a backup path-finding technique to protect the links against failures. Whenever a lightpath request arrives, there should be a backup to each link in the lightpath requests to protect the traversing demand. If there is a requirement of the backup path for some link, a cycle is formed around it, and assignment of slots is made, to be used when protection is activated. Considering that no two links fail simultaneously, if different links on the cycle require backup, then the slots can be shared. After the requests are released on completion of connection duration, the cycles if no more needed, are also dismantled. Since there is no pre-configuration of the cycles, i.e., cycles are formed on the fly, these cycles are called Dynamic Cycles (D-cycles).

\begin{figure}
\begin{subfigure}{0.5\textwidth}
    \centering
    \includegraphics[width=\linewidth]{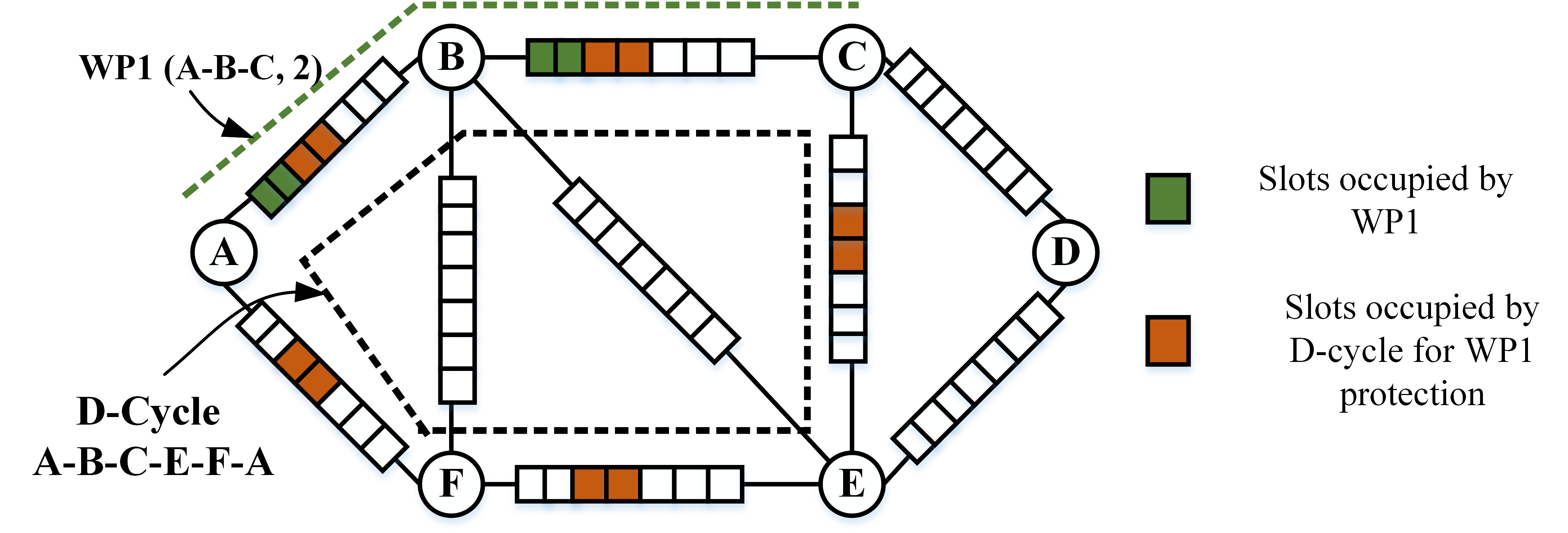}
    \caption{}
    \label{fig:dc1}
\end{subfigure}
\begin{subfigure}{0.5\textwidth}
    \centering
    \includegraphics[width=\linewidth]{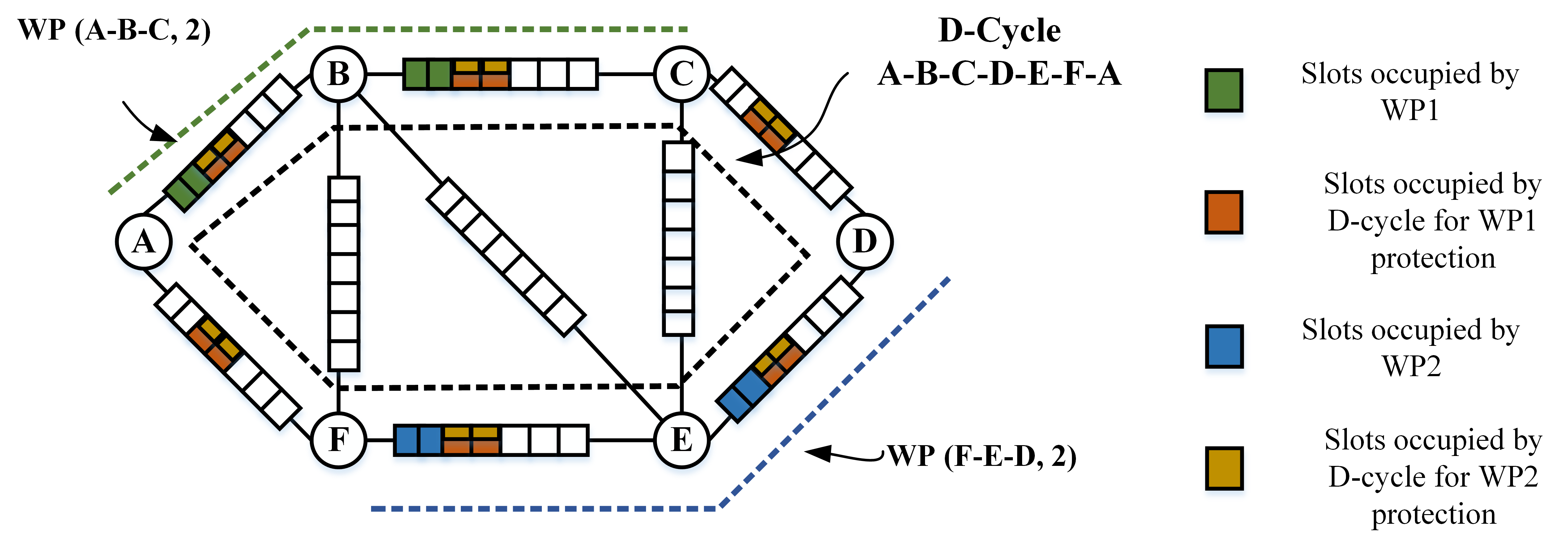}
    \caption{}
    \label{fig:dc2}
\end{subfigure}
\begin{subfigure}{0.5\textwidth}
    \centering
    \includegraphics[width=\linewidth]{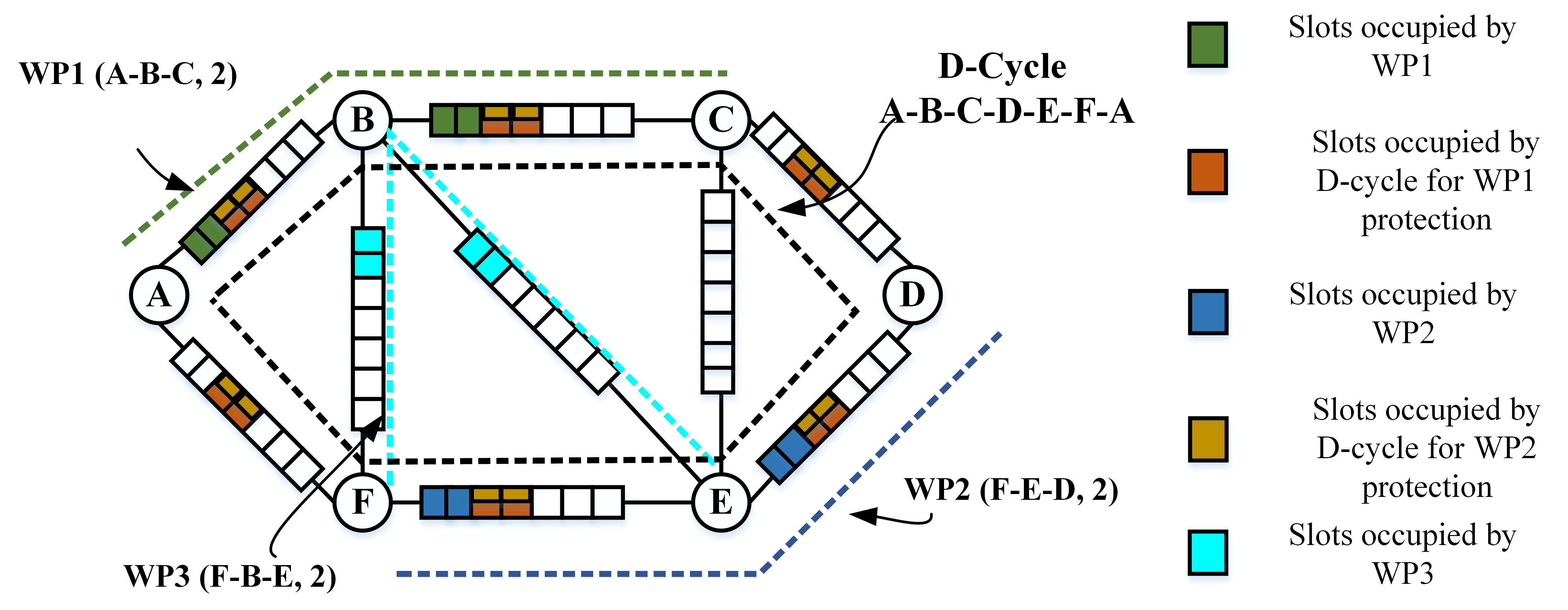}
    \caption{}
    \label{fig:dc3}
\end{subfigure}
\caption{Dynamic Cycles for three working paths.}
\label{fig:dcyc}
\end{figure}

Figure \ref{fig:dcyc} shows an example of Dynamic Cycles. There are three working lightpath requests A-B-C, F-E-D, and F-B-E. 

In figure \ref{fig:dc1}, Working Path WP1 (A-B-C) with two slots requirement is setup. Considering at max a single fault, a d-cycle A-B-C-E-F-A is created to protect Working Path, WP1. 

Another Working Path, WP2 (F-E-D) with two slots requirement is the next one created, as shown in figure \ref{fig:dc2}. Considering single fault, same d-cycle A-B-C-E-F-A can be used to protect link F-E of WP2. For link E-D either we can create second d-cycle or as we have done in the figure here, extend the existing d-cycle to A-B-C-D-E-F-A. Since, only single link needs protection in case of failure, backup slots used can be shared.

Working Path, WP3 (F-B-E) with two slots requirement, is the next one to be setup as shown in figure \ref{fig:dc3}. Same d-cycle A-B-C-E-F-A can be used to protect WP3. Here, link F-B and B-E are straddling links. Since, only single link needs protection in case of a failure, shared backup slots used on A-B-C-E-F-A can be used for protection of WP3 links. Here, two slots D-Cycle can be used to protect four slot working paths on the straddling links.

Here, each of the links in the requests is provisioned with a d-cycle. The spectrum slots are shared for different lightpath requests. Also, working links should be disjoint. As sharing of slots cannot be done if the working links are not disjoint. If the sharing of slots does not follow this constraint and a link fails, only one path can be protected; the other paths simply breaks.

\begin{itemize}
\item [\textbf{Pros}]
\item Incorporates the advantages of efficiency of mesh and protection speed of ring networks.   
\item The switching speed of link protection ($< 50 milliseconds$) is much faster than the path protection due to the faster fault localization.
\item Only the affected link on a path is provisioned with backup. All other links on the path remain intact.
\item As only one link is protected, there is no storm all across the network for setting up protection paths for all the paths passing through the failed link.
\item [\textbf{Cons}]
\item Costly spectrum converters (waveband shifters) are needed at the end nodes of the failed link, for restoring the paths.
\end{itemize}

The provisioning of a backup path for every link or path increases redundancy. In this thesis, we are using the Availability of an optical link and path to decide on if the costly protection should be provisioned or not, for a link. We discuss Reliability and Availability in the next section. 

\section{Reliability of an Optical Path}
The reliability of an optical path is the probability that all of its links are working without failure for an expected duration \cite{rb1}. The availability of each link is needed for the reliability analysis of a path. An optical path is a sequence of series-connected optical links and it will work if and only if all of its links are working. 

Consider a path \textit{p} consisting of \textit{l} links and suppose that each one of them is in either working or failed state. To indicate whether each link is working or not, we use a binary indicator variable $x_{i}^{p}$ as represented by equation \ref{eq:sv};

\begin{equation}
    x_{i}^{p} = \begin{cases}
    1, &\text{if $i^{th}$ link of path \textit{p} is working,}\\
    0, &\text{otherwise.}
\end{cases}
\label{eq:sv}
\end{equation}

The state vector ${x^{p}} = (x_{1}^{p}, x_{2}^{p},......, x_{l}^{p})$ indicates the state of the constituent links of the path \textit{p} i.e., either they are working or failed \cite{ra}. For lightpaths, whether or not the whole path is working or not is determined by the structure function $\pi_{p}(x^{p})$ of the path and indicated by equation \ref{eq:sf};

\begin{equation}
    \pi_{p}(x^{p})= \begin{cases}
    1, &\text{if the path \textit{p} is working when}\\
     & \text{the state vector is x,}\\
    0, &\text{otherwise.}
\end{cases}
\label{eq:sf}
\end{equation}

As we know the lightpath consists of series of independent links. Therefore, the structure function of series system is given by equation \ref{eq:sfs};

\begin{equation}
    \pi_{p}(x^{p})
    = \prod_{i=1}^{l} x_{i}^{p}.
\label{eq:sfs}
\end{equation}

\begin{figure}[h]
 \centering
    \includegraphics[width=0.5\linewidth]{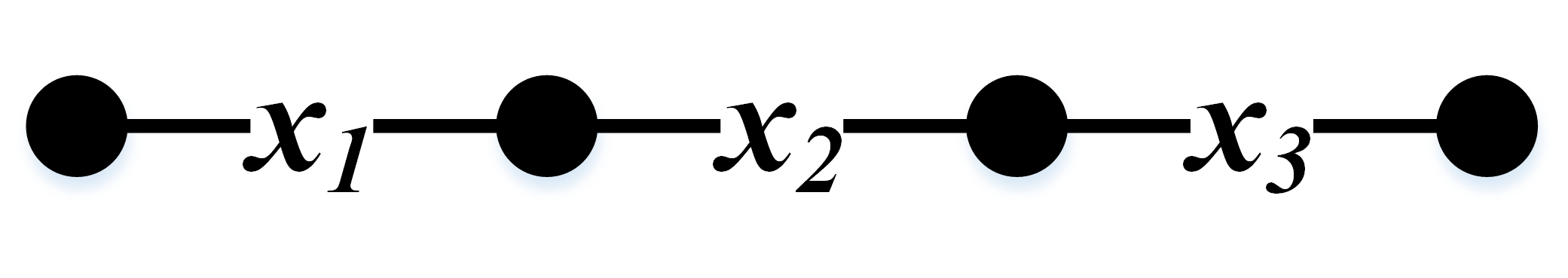}
    \caption{A four-nodes and three-links lightpath request where $x_1$, $x_2$, and $x_3$ are the binary variables representing whether corresponding links are working or not.}
    \label{fig:ex4n3l}
\end{figure}

Let us explain this with the help of an example. Consider a four-nodes and three-links lightpath as shown in figure \ref{fig:ex4n3l}. $\pi_{p}(x)$ for the lightpath is given in table \ref{table:states} using the above equations \cite{ra}. 

\begin{table}[h]	
    \begin{center}
        \centering        
        
        \begin{tabular}{|c|c|c|c|}
            \hline
            
            \textbf{$x_1$} & \textbf{$x_2$} & \textbf{$x_3$}& \textbf{$\pi(x)$} \\
            \hline
            0&  0 & 0 & 0 \\
            \hline
            0&  0 & 1 & 0 \\
            \hline
            0&  1 & 0 & 0 \\
            \hline
             0&  1 & 1 & 0 \\
            \hline
             1&  0 & 0 & 0 \\
            \hline
             1&  0 & 1 & 0 \\
            \hline
             1&  1 & 0 & 0 \\
            \hline
             1&  1 & 1 & 1 \\
            \hline
             
        \end{tabular}
        \caption{Different states of the links of an optical path with four-nodes and three-links, where 1 represents that the link is working and 0 represents that the link is in the failed state.}
        \label{table:states}
    \end{center}    
\end{table}

Let us suppose $X_{i}$ be a random variable that represents the state of the $i^{th}$ link.

\begin{equation}
    P \{X_{i} = 1 \} = p_i = 1-P \{X_{i} = 0 \}.
\label{eq:r1}
\end{equation}

$p_i$ in equation \ref{eq:r1} is the probability that the $i^{th}$ link is working, and is defined as the Reliability of the $i^{th}$ link. Therefore, reliability function \textit{r(p)} of an optical path \textit{p} with series of connected independent links is given by

\begin{equation}
    r(p) = P \{\pi_{p}(x^{p}) = 1 \}.
\label{eq:r2}
\end{equation}
 
\begin{figure}[h]
 \centering
    \includegraphics[width=\linewidth]{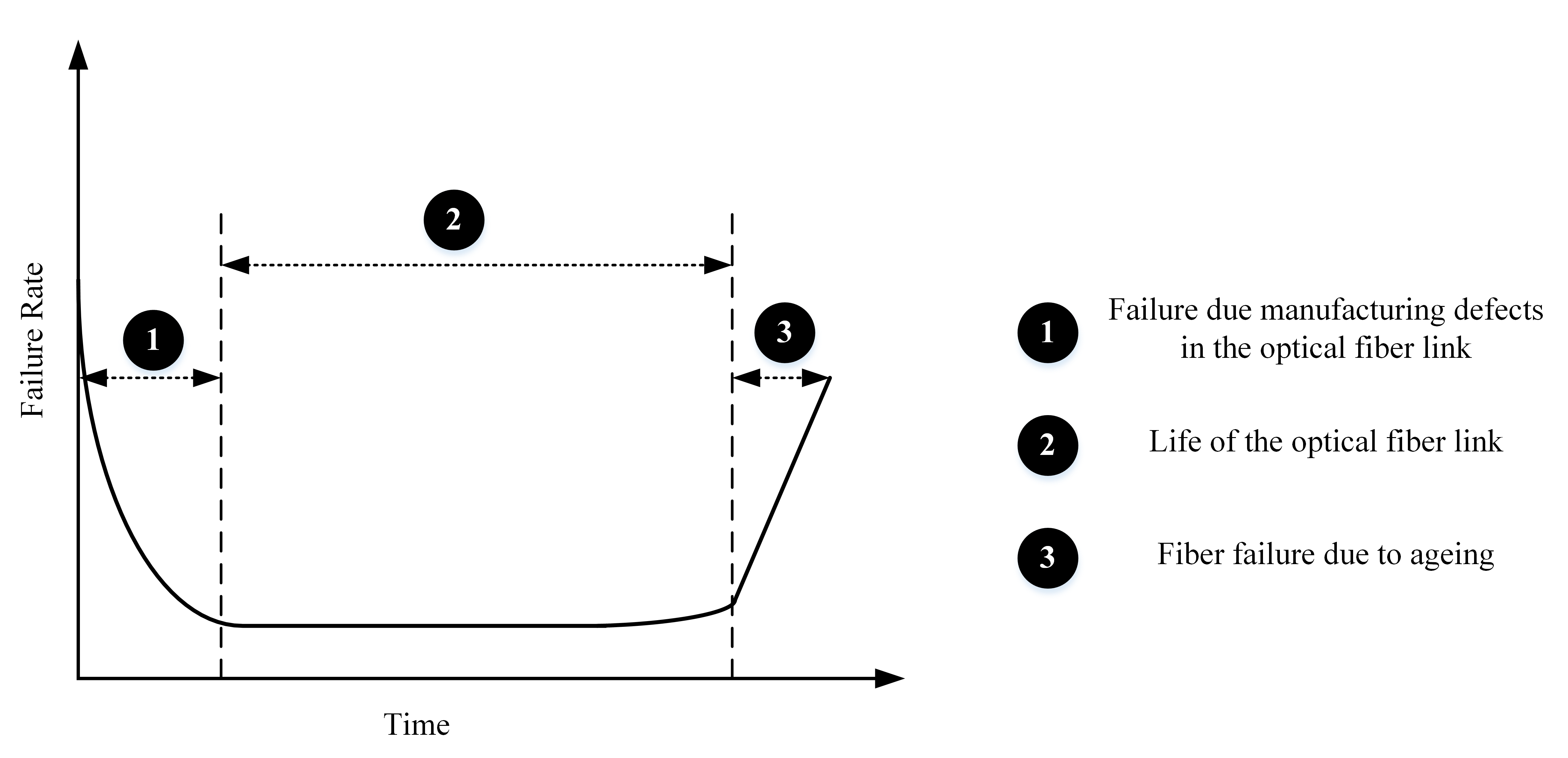}
    \caption{Bathtub curve \cite{bath} representing the failure rate in an optical fiber link with time.}
    \label{fig:bathtub}
\end{figure}

Reliability of an optical fiber link can be modeled with the help of Bathtub curve as shown in figure \ref{fig:bathtub}. It maybe noted that the reliability is computed for non-repairable systems. 

The optical fiber links are repairable components. Therefore, in this thesis, we consider Availability of an optical fiber instead of reliability, which takes into account both failures and repairs.

\section{Availability Analysis of an Optical Path}

Consider a path with \textit{l} links having reliability $r(p)$. Let us suppose, all the links are working initially.

\begin{equation}
    A(t) =  P \{path \ is \ working \ at \ time \ t \}
\label{eq:ava1}
\end{equation}


\textit{A(t)} is the availability of an optical path at time \textit{t}, and $A_{e}(t)$ is the availability of $e^{th}$ link at time \textit{t} \cite{ra}. Since we are considering repairable link, so a link can be in two states i.e., under repair after failure and normally operating. 

Availability of an optical path can be represented as

\begin{equation}
    A_{p}(t) =  \prod_{e \in P} A_{e}(t), 
\label{eq:ava3}
\end{equation}

 where \textit{p} is the lightpath request and \textit{e} is the link belonging to path \textit{P}. 
 
\begin{figure}[h]
 \centering
    \includegraphics[width=\linewidth]{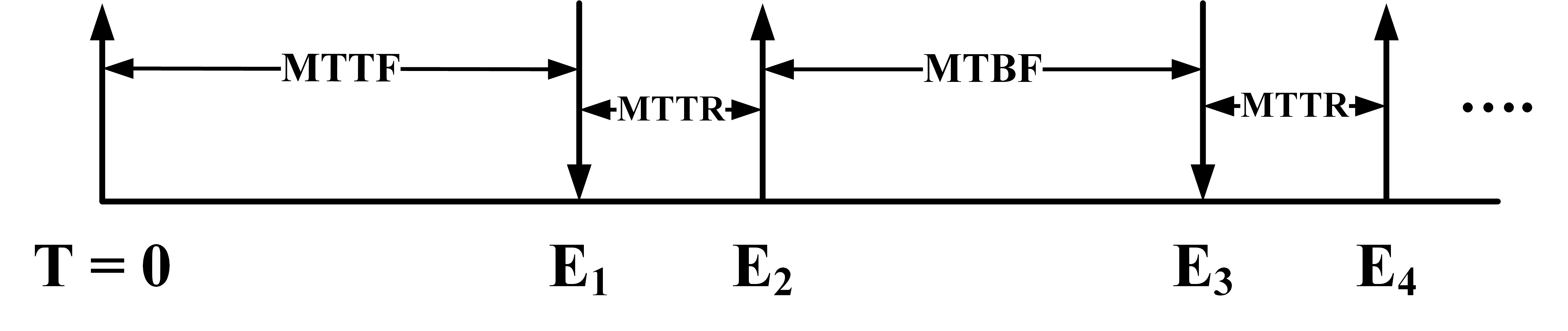}
    \caption{Availability analysis of a link.}
    \label{fig:AA}
\end{figure}

In figure \ref{fig:AA}, MTTF is the Mean Time To Failure, MTTR is the Mean Time To Repair and MTBF is the Mean Time Between Failures. $E_1$ is the event when first failure occurs. After a certain repair time, event of link getting repaired $E_2$ happens and next failure happens as event $E_3$ and so on.

Suppose that the link works for an exponentially distributed time with rate $\lambda' = \dfrac{1}{MTTF} = \dfrac{1}{MTBF}$ and then fails. Once failed, it takes an exponentially distributed time with rate $\mu' = \dfrac{1}{MTTR}$ to be repaired. Therefore, Availability of an optical link $e$ is

\begin{equation}
    A_{e}(t) =  \dfrac{\ MTTF}{\ MTTF + \ MTTR}, \text{ or}
\label{eq:ava4}
\end{equation}

\begin{equation}
    A_{e}(t) =  \dfrac{\ MTBF}{\ MTBF + \ MTTR}.
\label{eq:ava5}
\end{equation}

MTTF (MTBF) is the average time of link operations since start (last repair) to next failure. MTTR is the average time needed to repair the link.

\subsection{Availability of Different Systems}

In this section we will show how to compute the Availability of different systems \cite{ra}, assuming each link to be independent.

\subsubsection{Series System}
An optical lightpath is series-connected optical links or edges, where a link $e \in E$; \textit{E} is the set of edges in the optical network.
\begin{figure}[h]
 \centering
    \includegraphics[width=\linewidth]{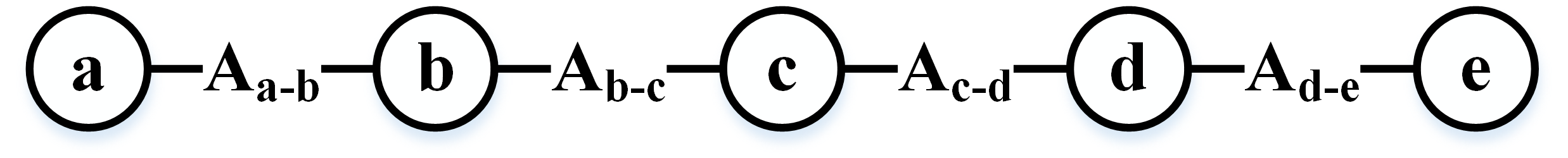}
    \caption{Series-connected optical path.}.
    \label{fig:series}
\end{figure}

Figure \ref{fig:series} shows series-connected links forming an optical lightpath. A five nodes four links lightpath has Availability of each link as $A_{a-b}, A_{b-c}, A_{c-d}, $ and $A_{d-e}$. The Availability of the series-connected path is given by

\begin{equation}
    A_{series} =  \prod_{e \in P} A_{e}, 
\label{eq:s1}
\end{equation}

where $A_{series}$ is the Availability of a series-connected path, and $A_e$ is the Availability of a link \textit{e} belonging to path \textit{P}. Thus, the Availability of path in figure \ref{fig:series}, is given by 

\begin{equation}
    A_{series} =  A_{a-b} A_{b-c} A_{c-d} A_{d-e}.
\label{eq:s2}
\end{equation}

The above is an example of series-connected system of working lightpath with no backup.

\subsubsection{Parallel System}
\begin{figure}[h]
 \centering
    \includegraphics[width=\linewidth]{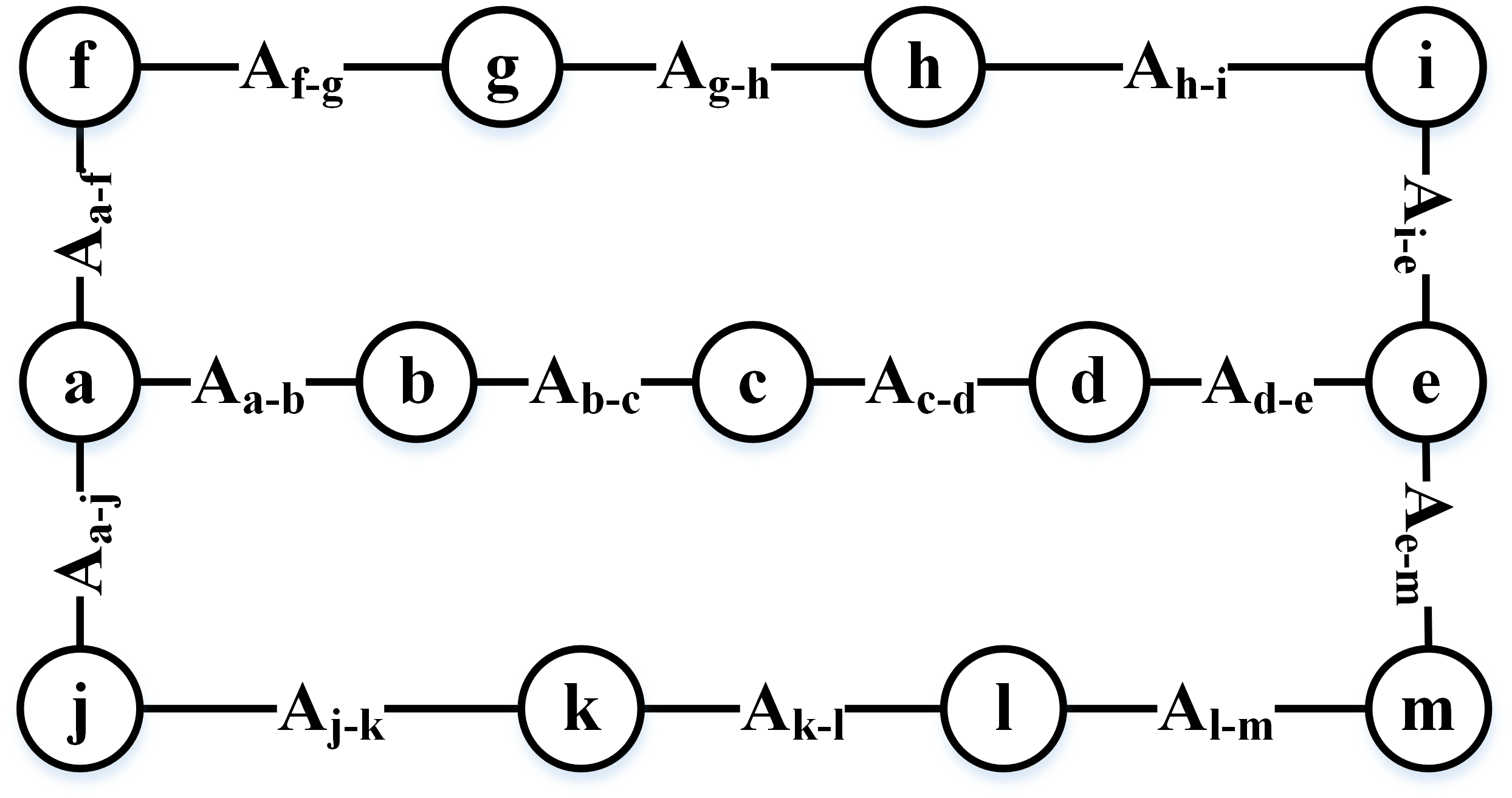}
    \caption{Parallel connected optical paths from \textit{a to e}.}
    \label{fig:parallel}
\end{figure}

In parallel system, redundant paths are found between source-destination pairs. In other words, the connection of multiple series paths for a single source-destination pair forms a parallel system (figure \ref{fig:parallel}). 


\begin{equation}
    A_{parallel} =   1-(1-\prod_{e \in P_l}A_{e})^{l}, 
\label{eq:p1}
\end{equation}

where $A_{parallel}$ is the Availability of a path between source and destination through parallel connected provisioned capacity, $A_e$ is the Availability of a $e^{th}$ link belonging to path $P_l$ where \textit{l} is the number of redundant paths. 

\begin{equation}
    A_{p'} =  A_{a-b}A_{b-c}A_{c-d}A_{d-e} 
\label{eq:p2}
\end{equation}

\begin{equation}
    A_{p''} =  A_{a-f}A_{f-g}A_{g-h}A_{h-i}A_{i-e}
\label{eq:p3}
\end{equation}

\begin{equation}
    A_{p'''} =  A_{a-j}A_{j-k}A_{k-l}A_{l-m}A_{e-m} 
\label{eq:p4}
\end{equation}

Here, $A_{p'}$, $A_{p''}$, and $A_{p'''}$ are the Availability of a three parallel connected paths. Availability of a parallel system as shown in figure \ref{fig:parallel} is given by 

\begin{equation}
    A_{parallel} =  1-((1-A_{p'})(1-A_{p''})(1-A_{p'''})) 
\label{eq:p5}
\end{equation}

An example of a parallel connected system is Shared Backup Path Protection, where for a working path, multiple backup paths can be provisioned depending upon the number of simultaneous failures expected. In the example of figure \ref{fig:parallel}, double faults can be tolerated, i.e., one fault on the working path and another one on the backup path. Third path can be used as a backup for the working path in case of two faults. Due to redundancy, $A_{parallel} > A_{series}$.

\subsubsection{Series-Parallel System}

In series-parallel system, the backup paths for series connected links are computed only for the affected (failed) links. In other words, a series system is converted into series-parallel when there are faulty links. Additional resources needed in such systems are less as compared to parallel systems. 

\begin{figure}[h]
 \centering
    \includegraphics[width=\linewidth]{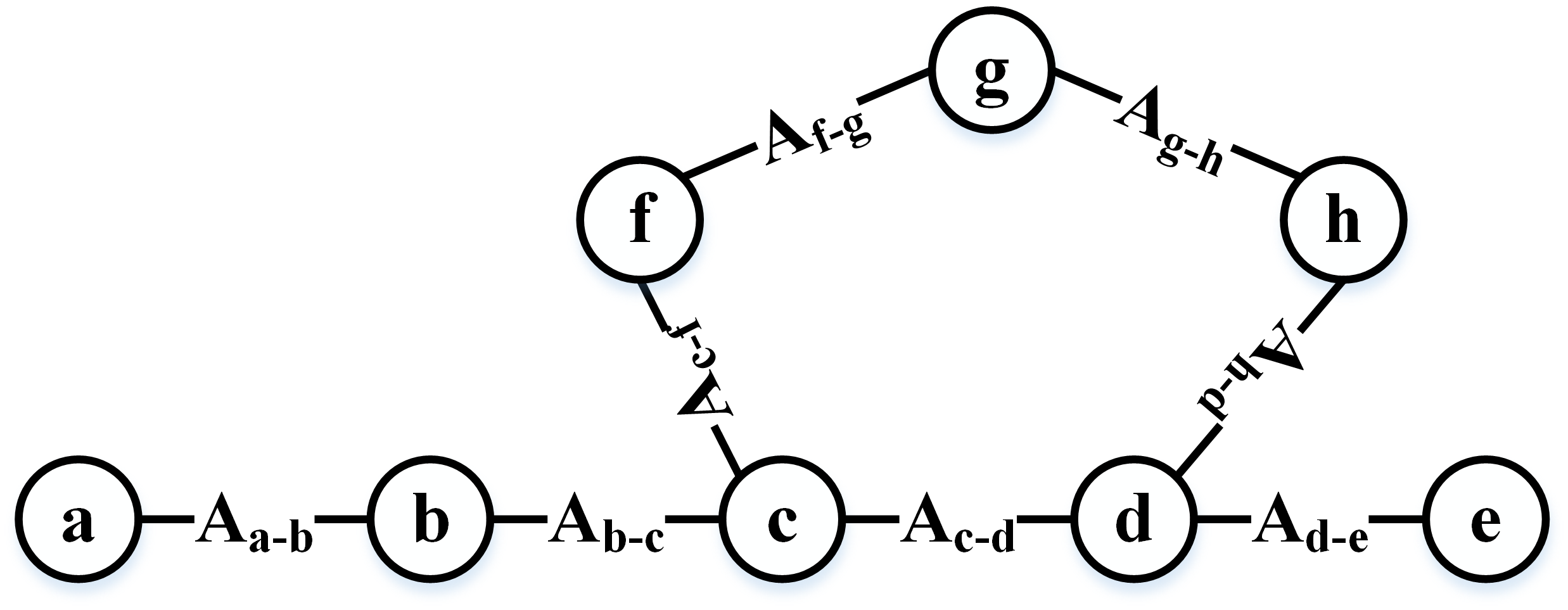}
    \caption{Series-connected optical path with Parallel-connected backup path for an optical edge $c-d$.}
    \label{fig:sp}
\end{figure}

%

\begin{equation}
    A_{series-parallel} = \prod_{e \in k}[1-((1-A_e)(1-A'_{e}))] \prod_{e \in L-k} A_e, 
\label{eq:sp1}
\end{equation}
where \textit{k} is the set of links which can go faulty. \textit{L} is the set of links in the working path, $A_e$ is the Availability of link \textit{e}, and $A'_{e}$ is the Availability of protection path for link \textit{e}.

\begin{multline}
    A_{series-parallel} =  A_{a-b}A_{b-c}(1-(1-A_{c-d})(1-(A_{c-f} \\ 
     A_{f-g}A_{g-h}A_{h-d})))A_{d-e}.
\label{eq:sp2}
\end{multline}
 
Equation \ref{eq:sp2} is the Availability of series-parallel system for figure \ref{fig:sp}. In this example, out of four links, one link with low availability is provided with backup path. 

Examples of series-parallel systems are working paths with p-cycles or d-cycles protecting the fault prone links. 

\section{Proposed Algorithms}

In this paper, we propose two algorithms based on the availability of the links and paths. For Routing and Spectrum Assignment of a Working path, we use the algorithm from \cite{DRSA}, i.e., Type II: Routing and Spectrum Assignment based on Consecutive Spectrum Slots. For a Backup path, static parameters such as hops or distance can increase the time complexity within the network. For example, if we find the alternate backup path using the shortest path algorithm, sometimes the required resources may not be available. Finding the required resources becomes more challenging as the number of incoming requests per unit time (or load in the network) increases. Instead of the shortest path algorithm for the backup path, we use the Type II Algorithm for the backup path also. 

\subsection{Notations Used in the Algorithms}
\begin{itemize}
    \item $G(V,E, \{ \Delta_{e} \})$: represents an optical network as a graph.
    \newline
    Here,
    \begin{itemize}[label={}] 
        \item $V$ is the set of nodes represented as vertices of the graph, indexed by \textit{v},
        \item $E$ is the set of optical fiber link represented as edges of the graph, indexed by \textit{e}, and
        \item $\Delta_{e}$ is the sequence of 1s and 0s  to model the availability status of the spectrum slots on the edge $e \in E$.
    \end{itemize}
    \item $LR(s,d,\{ \Delta^{r} \}, k)$ is a Lightpath Request where
    \begin{itemize}[label={}]
        \item $s$ is the source node, $s \in V$,
        \item $d$ is the destination node, $d \in V$ ,
        \item $ \Delta^{r} $ is the bitmap of required contiguous and continuous spectrum slots from $s$ to $d$, and
        \item $k$ is a positive integer. It is the maximum number of paths to be computed by RSA. 
    \end{itemize}
    \item $\Delta_{p}$ is the bitmap of available spectrum slots in path $p$ from $s$ to $d$.
    \item $(u, v)$ is the edge joining the pair of vertices (\textit{u} and \textit{v}), where $u$ is the starting (head) node and $v$ is the ending (tail) node.
    \item $A_{p}^{max}$ is the Maximum Availability of the working path selected from the set of candidate working paths.
    \item $A_{pp}^{max}$ is the Maximum Availability of the working path that needs protection.
    \item $A_{bp}^{max}$ is the Maximum Availability of the backup path selected from the set of candidate backup paths.
    \item $A_{th}$ is the Threshold Availability.

\end{itemize}
Algorithm \ref{alg:type2b} is the Routing and Spectrum Assignment with Consecutive Slots (RSACS) algorithm used for finding both working paths as well as backup paths. In this algorithm, \textsc{CandidatePaths()} function (Algorithm \ref{fun:cpb}) is used to find set of candidate working paths from source node to the destination node. It also returns the value of the Availability $A_p$ of each candidate path $p$. The function of \textsc{CandidatePaths()} has already been explained in paper (DRSA). 

In the algorithm RSACS with Protection provisioning based on the Availability of the path, we select the path with maximum availability $A_{p}^{max}$. The backup path provisioning is done based on the Availability of the path $A_{p}$. The decision of whether to consider a path with protection provisioning depends on the value of the Availability Threshold, $A_{th}$. The protection is provisioned for the paths whose Availability $A_{p}^{max}$ value is less than the threshold value. Then the notation of $A_{p}^{max}$ changes to $A_{pp}^{max}$, i.e., the Maximum Availability of the working path that needs protection. Depending on the protection type, the backup paths are provided to the less available working paths or links. 

We consider two types of protection. First is Dynamic Shared Backup Path and Spectrum Slots (DSBPSS) are used for path protection with slot sharing. Another one is Dynamic Cycles (modified p-Cycles), used for link protection with slot sharing.

\begin{algorithm*}
    \caption{RSACS with Protection provisioning based on Availability of the path}
    \begin{algorithmic}[1]
        \State \textbf{Input:} $G(V, E, \{ \Delta_{e} \}, A_e )$, $LR(s, d,  |\Delta^{r}|, k), A_{th}, BackupPaths, DCycles$
        \State $ AllPath \leftarrow$ A set of candidate paths from $s$ to $d$ using \Call{CandidatePaths}{$G(V, E, \{ \Delta_{e} \}, A_e)$, $LR(s, d, |\Delta^{r}|, k)$} and the Availability $A_p$ of the candidate paths \textit{p} \Comment{\textcolor{blue}{Algorithm \ref{fun:cpb}}}
         \If {$AllPath$ is not empty}
         	
         		\State Select the path with highest Availability $A_{p}^{max}$ from all then paths listed in \textit{AllPath} and store in \textit{BestPath} 
            
            \If {$A_{p}^{max} < A_{th}$} 
            	\State $A_{pp}^{max} = A_{p}^{max}$
            	\If {Protection type is DSBPSS}          
            	  	
            	\State $BackupPath$ =  \Call{DSBPSS}{$G(V, E, \{ \Delta_{e} \}, A_e), LR(s, d, |\Delta^{r}|, k),BestPath, $\par
        \hskip\algorithmicindent $BackupPath, A_{pp}^{max}, A_{th}$}  \Comment{\textcolor{blue}{\Call{DSBPSS}{} function is used for path protection}}
        		\If {$BackupPath$ is not empty}
        			\State $BackupPaths = BackupPaths + BackupPath$
            		\State \textbf{Return} $BestPath$, $BackupPaths$
            		\Else 
            		\State \textbf{Return} $BestPath$
            	\EndIf
            	\Else
            		
            		\State $Cycles$ =  \Call{Dcyc}{$G(V, E, \{ \Delta_{e} \},A_e), LR(s, d, |\Delta^{r}|, k),$\par
        \hskip\algorithmicindent $ BestPath, A_{pp}^{max}, A_{th}, Cycles$} \Comment{\textcolor{blue}{\Call{Dcyc}{} function is used for link protection. Here, list of links which are protected and updated cycles (each cycle is list of links) should be returned, such that $A_{pp}^{max} \geq A_{th}$}}
        \If {$Cycles$ is not empty}
        			
        			\State $DCycles = DCycles + Cycles $
            		\State \textbf{Return} $BestPath$, $DCycles$
            		\Else 
            		\State \textbf{Return} $BestPath$
            	\EndIf
            		
            	\EndIf         	
            \Else
            	\State \textbf{Return} $BestPath$
            \EndIf               
         \Else
         	\State Block the request
         \EndIf
        
    \end{algorithmic}
    \label{alg:type2b}
\end{algorithm*}

\begin{algorithm}
\caption{\textsc{CandidatePaths()} $\rightarrow$ Path with contiguous slots $\geq$ $\Delta^r$}
    \begin{algorithmic}[1]
        \Function{CandidatePaths}{$G(V, E, \{ \Delta_{e}\}, A_e )$, $LR(s, d, |\Delta^{r}|)$, $k$}
       
        \State $AllPath$ = [ ] \Comment{\textcolor{blue}{$AllPath$ contain paths and their availability.}}
        \State $Path = [s, 1]$ \Comment{\textcolor{blue}{Element of Path and its availability.}}
        
        \While {$Path$ is not empty}
            \State $tmp$ = [ ]
            \For {all paths in \textit{Path} indexed by $i$}
                \State $u = Path(i).end$
                \State $\{v\} = \Call{Adj}{u}) \backslash Path(i)$ \Comment{\textcolor{blue}{All neighbours of \textit{u} which are already not there on the \textit{Path(i)}.}}
                \If {$\{v\} \neq $ [ ]}
                    \For {all nodes in $\{ v \}$ indexed by $j$}
                        \State $\Delta_{i} \leftarrow \Delta_{i} \cap \Delta_{(u,v(j))}$
                        \If {\Call{IsFeasible}{$(\Delta_i, |\Delta^{r}|$} $== \textsc{True}$}
                                \If {$v(j) == d $}
                                    
                                    \State $A_p = A_p \times A_{(u,v(j))}$
                                    \State $AllPath = AllPath + [(Path(i), v(j)); A_p]$
                                    \If {$size(AllPath) == k$}
                                        \State \textbf{Return:} $AllPath$
                                    \EndIf
                                    
                                \Else
                                	\State $A_p = A_p \times A_{(u,v(j))}$
                                    \State $tmp = tmp + [(Path(i), v(j)); A_p]$
                                    
                                \EndIf
                            
                            \EndIf
                    \EndFor
                \EndIf
            \EndFor
            \State $Path = tmp$
        \EndWhile
        \State \textbf{Return:} $AllPath$ 
        \EndFunction
    \end{algorithmic}
    \label{fun:cpb}
\end{algorithm}

\begin{algorithm}
\caption{\textsc{IsFeasible()}}
    \begin{algorithmic}[1]
    \Function {IsFeasible}{$(\Delta_a, |\Delta_{b}|)$}
        \Comment{\textcolor{blue}{\Call{IsFeasible}{} function checks the $|\Delta_{b}|$ number of contiguous slots are available in the bitmap $\Delta_a$}}
        \If {$|\Delta_{b}|$ can be accommodated in $\Delta_a$}
            \State \textbf{Return:} \textsc{True}
        \Else
        	\State \textbf{Return:} \textsc{False}
        \EndIf
    \EndFunction
    \end{algorithmic}
    \label{fun:ifb}
\end{algorithm}

\subsection{RSACS with DSBPSS}

\begin{algorithm*}
\caption{\textsc{DSBPSS()}}
    \begin{algorithmic}[1]
    \Function {DSBPSS}{$G(V, E, \{ \Delta_{e} \}, A_e), LR(s, d, |\Delta^{r}|, k), BestPath, BackupPathList, $\par
        \hskip\algorithmicindent$A_{pp}^{max}, A_{th}$}    

 		\State $BackupPath$ = [ ] 	\Comment{\textcolor{blue}{$BackupPath$ stores the backup path values}}
 		\State $ G(V', E', \{ \Delta'_{e} \}, A'_e) = \Call{RemoveLinks}{G(V, E, \{ \Delta_{e} \}, A_e), BestPath}$ \Comment{\textcolor{blue}{$\Call{RemoveLinks}{}$ function remove links of \textit{BestPath} from $G(V, E, \{ \Delta_{e} \}, A_e)$. This is done for finding working links disjoint backup paths }}
 		\State $G(V', E', \{ \Delta'_{e} \}, A'_e) = \Call{FBS}{G(V', E', \{ \Delta'_{e} \}, A'_e), BackupPathList}$ \Comment{\textcolor{blue}{$\Call{FBS}{}$ function find all locally free slots for backup path. These slots may have been already allocated to other backup paths in \textit{BackupPaths} in $G(V', E', \{ \Delta'_{e} \}, A'_e)$. This is done for slots sharing between backup links.}}
 		\State $  BackupPaths \leftarrow$ A set of candidate paths from $s$ to $d$ using \Call{CandidatePaths}{$G(V', E', \{ \Delta'_{e} \}, A_e)$, $LR(s, d, |\Delta^{r}|, k)$} and their corresponding Availability $A_{bp}$ \Comment{\textcolor{blue}{Algorithm \ref{fun:cpb}}}
        \While {$A_{pp}^{max} < A_{th}$}   \Comment{\textcolor{blue}{Here $A_{pp}^{max}$ is the availability of the protected path. }}    	
        	\If {$BackupPaths$ is not empty}        	
         		\State Remove the path with highest Availability from all paths listed in \textit{BackupPaths} and store in \textit{BackupPath} with its corresponding $A_{bp}^{max}$ \Comment{\textcolor{blue}{$A_{bp}^{max}$  is the availability of the backup path}}         		
         	\Else         	
         		\State \textbf{Return:} \textit{BackupPath} = [ ]
         	\EndIf  	
 		\State $A_{pp}^{max} = \Call{AvaDSBPSS}{A_{pp}^{max}, A_{bp}^{max}}$
 	       		
        \EndWhile
        \State \textbf{Return:} \textit{BackupPath}
    \EndFunction
    \end{algorithmic}
    \label{fun:sbrp}
\end{algorithm*}
In Algorithm \ref{alg:type2b} line number 8, if the protection type is DSBPSS, the function in Algorithm \ref{fun:sbrp} is used for finding path protection. The remaining candidate paths in \textit{AllPath} (Algorithm \ref{alg:type2b}) cannot be used for backup path, since slots sharing is not feasible as we want to find slots disjoint candidate paths in $AllPath$ (Algorithm \ref{alg:type2b}). 

In \textsc{DSBPSS()} function, first \textsc{RemoveLinks} function is used to remove the links from the graph. For finding link disjoint backup path, we first exclude the current working path's links from the graph locally. Now, for corresponding backup path all locally free slots which may also have been allocated to all the other backup paths are found using Free Backup Slots (FBS) function. It allows already allocated backup slots also to be shared. Next, for finding backup paths and their corresponding Availability, \textsc{CandidatePaths()} function is used. The backup path with higher $A_{bp}^{max}$ is selected for the backup path. 

The backup paths availability is updated with the help of function as in Algorithm \ref{fun:asrbp}. After updating the $A_{pp}^{max}$ value again, it is verified to be greater than $A_{th}$ or not. If not, then from the list of backup paths (excluding already selected backup paths), we select another path with maximum Availability. Therefore, the working path is provisioned with redundant backup paths until the $A_{pp}^{max} \geq A_{th}$. 
\begin{algorithm}
\caption{\textsc{AvaDSBPSS()}}
    \begin{algorithmic}[1]
    \Function {AvaDSBPSS}{$ A_{pp}^{max}, A_{bp}^{max}$}
    	
        \State $A_{pp}^{max} = 1-((1-A_{pp}^{max})\times(1-A_{bp}^{max}))$ \Comment{\textcolor{blue}{$A_{pp}^{max}$ is updated after backup path provisioning.}}
    \EndFunction
    \end{algorithmic}
    \label{fun:asrbp}
\end{algorithm}

\subsection{RSACS with D-cycles}

In Algorithm \ref{alg:type2b} line number 8, if the protection type is \textsc{Dcyc}, the function as in Algorithm \ref{fun:dcyc} is used for link protection.  A set of working links $(l_1, l_2, l_3,...)$ forms a working path (\textit{p}).

\begin{algorithm*}
\caption{\textsc{Dcyc()}}
    \begin{algorithmic}[1]
    \Function {Dcyc}{$G(V, E, \{ \Delta_{e} \},A_e), LR(s, d, |\Delta^{r}|, k), BestPath, A_{pp}^{max}, A_{th}, Cycles$} 
    	\State Store in $A$ the value of Availability of the links $A_{e}$ used \Comment{\textcolor{blue}{$ A$ store the values of the Availability of the links that form the working path.}} 
    	\State  $BackupLinks$ = [ ]		
        \While {$A_{pp}^{max} < A_{th}$} 
            
        	\State Find $A_{l}$ = \Call {$Min_{l}$}{\textit{A}} \Comment{\textcolor{blue}{$A_{l}$ is the Availability of the minimum available link \textit{l}.}}       	
        	
        	\If {\Call{CheckCycles}{$Cycles, LR(s, d, |\Delta^{r}|, l)$} == True} \Comment{\textcolor{blue}{\Call{CheckCycles}{} checks an existing cycle in \textit{Cycles} can be used for protection of link \textit{l}.}}
        		
        		\State $[Cycles, A^{max}_{bp}]$ = \Call{UpdateCycles}{$Cycles, |\Delta^{r}|, l$}
        	\ElsIf {path $p'_1$ alternate to \textit{l} can be found} \Comment{\textcolor{blue}{Algorithm \ref{fun:cpb}.}}
        		\If{check $p'_1$ can be used to extend existing cycles}
            		\State $[Cycles, A^{max}_{bp}]$ = \Call{UpdateCycles}{$Cycles, |\Delta^{r}|, l, p'_1$} 
            	\ElsIf{path $p'_2$ can be found as alternative to \textit{l}, while being node and link disjoint to $p'_1$ except to the end nodes of \textit{l}} \Comment{\textcolor{blue}{Algorithm \ref{fun:cpb}.}}
            		\State Use $p'_1$ and $p'_2$ to form new cycles \Comment{\textcolor{blue}{$p'_1$ and $p'_2$ must be link disjoint.}}
            		\State $[Cycles, A^{max}_{bp}]$ = \Call{UpdateCycles}{$Cycles, |\Delta^{r}|, l, p'_1, p'_2$}
            	\ElsIf{extra slots are there on link \textit{l}, to form a cycle with \textit{l} as oncycle}
            		\State $[Cycles, A^{max}_{bp}]$ = \Call{UpdateCycles}{$Cycles, |\Delta^{r}|, l, p'_1$}       	
            	
            	\Else
            		\State \textbf{Return:} No cycles found	
            		
            	\EndIf
            \Else
            	\State \textbf{Return:} No cycles found
           \EndIf
 	    \State $A^{max}_{pp}, A_{pl}  = \Call{AvaDcyc}{A^{max}_{pp}, A_{l}, A^{max}_{bp}}$ \Comment{\textcolor{blue}{$A_{pl}$ is the Availability of the protected link. Availability of the least available link which is modified by provisioning of the backup path to it.}}
 	   
 	    \State $A = (A-A_{l})+A_{pl}$ 	
 	       		
        \EndWhile
    \State \textbf{Return:} \textit{Cycles}  
    \EndFunction
    \end{algorithmic}
    \label{fun:dcyc}
\end{algorithm*}

In this method, the working link with minimum Availability (say $i^{th}$ link, $l_i \in p$) is selected from the set of working links for protection provisioning. For protection, cycles are formed using available capacity in the network. We used  \textsc{CheckCycles()} function to check whether protection is feasible or not using existing cycles in \textit{Cycles}. If no backup cycles found in \textit{Cycles}, then new cycles is formed for protection and stored in \textit{Cycles} along with their assigned slots.  

New cycles can provide either on-cycle or straddling protection. For on-cycles protection, set of backup paths are found using \textsc{CandidatePaths()} function for least available link, $l$. It also returns the Availability of the backup path $A_{bp}$. The path $p'_1$  with higher $A_{bp}$ is selected for the backup path. Out of three conditions, one should be satisfied as mentioned below.

\begin{itemize}

\item Check if $p'_1$ can be used to extend existing cycles from \textit{Cycles}, then update \textit{Cycles} and assign it for protection. \\

\item Find another path $p'_2$ path as an alternative to link $l$, and it should be node and link disjoint to $p'_1$ except the end nodes of $l$, such that it forms a straddling cycle with link $l$.  Update \textit{Cycles} and assign it for protection. \\

\item If any of the above two conditions are not satisfied, then check if extra slots are there on link $l$, to form a cycle with $l$ for on-cycle protection, then update \textit{Cycles} and assign it for protection. \\
\end{itemize}
\begin{algorithm*}
\caption{\textsc{AvaDcyc()}}
    \begin{algorithmic}[1]
    \Function {AvaDcyc}{$A^{max}_{pp}, A_{l}, A^{max}_{bp}$}
        	\State $A_{pl} = 1-((1-A_{l})\times (1-A^{max}_{bp}))$ 
        	\State $A^{max}_{pp} = \dfrac{A^{max}_{pp} \times A_{pl}}{A_{l}} $
        	\smallskip
        	\State \textbf{Return:} $A^{max}_{pp}, A_{pl}$
    \EndFunction
    \end{algorithmic}
    \label{fun:adcyc}
\end{algorithm*}

The backup paths availability is updated with the help of function in Algorithm \ref{fun:adcyc}. After updating the $A_{pp}^{max}$ value again, check either it is greater than $A_{th}$ or not. If not, then repeat the same procedure till the working path is provisioned with sufficient redundant backup paths to achieve $A_{pp}^{max} \geq A_{th}$. 

If none of the conditions are satisfied to form a backup cycle, then the working path is not protected.

\section{Numerical Results}

In paper [DRSA], the algorithm that outperforms every other is Type II: Routing and Spectrum Assignment based on Consecutive Slots. Therefore, we proposed algorithms in which the working paths and backup paths are found on the basis of contiguous and continuous spectrum available. The protection is provisioned to only those paths or links which have lower availability than the threshold values.

\subsection{Performance Metrics} 
The performance of the proposed algorithms is evaluated on the basis of blocking of the incoming requests, blocking of the incoming required slots, the spectrum utilization, Capacity Used for Protection and Restorability Ratio with the gradual increments of demand rate (in Gbps): 
    \begin{itemize}
        \item \textbf{Blocking Probability}: It is defined as the ratio of total number of blocked connections to the total number of arrived connections.
        \item \textbf{Bandwidth Blocking Probability}: It is defined as the ratio of the total amount of incoming bandwidth or slots blocked to the total amount of bandwidth or slots required for all the connections.
        \item \textbf{Spectrum Utilization}: The ratio of total bandwidth or slots used to the total bandwidth or slots in the spectrum.
        \item \textbf{Capacity Used for Protection}: It is the total amount of usable bandwidth used for protection\footnote{only for the paths whose Availability of the link or path is less than the Threshold Availability.
}. 
        \item \textbf{Restorability Ratio}: It is the ratio of total number of protected paths to the total number paths requiring protection (sum of protected path and unprotected paths). This ratio is only for the paths whose Availability is less than the Threshold Availability\footnote{only for the paths whose Availability of the link or path is less than the Threshold Availability.
}.

    \end{itemize}
\begin{figure}
    \centering
    \includegraphics[width=\linewidth]{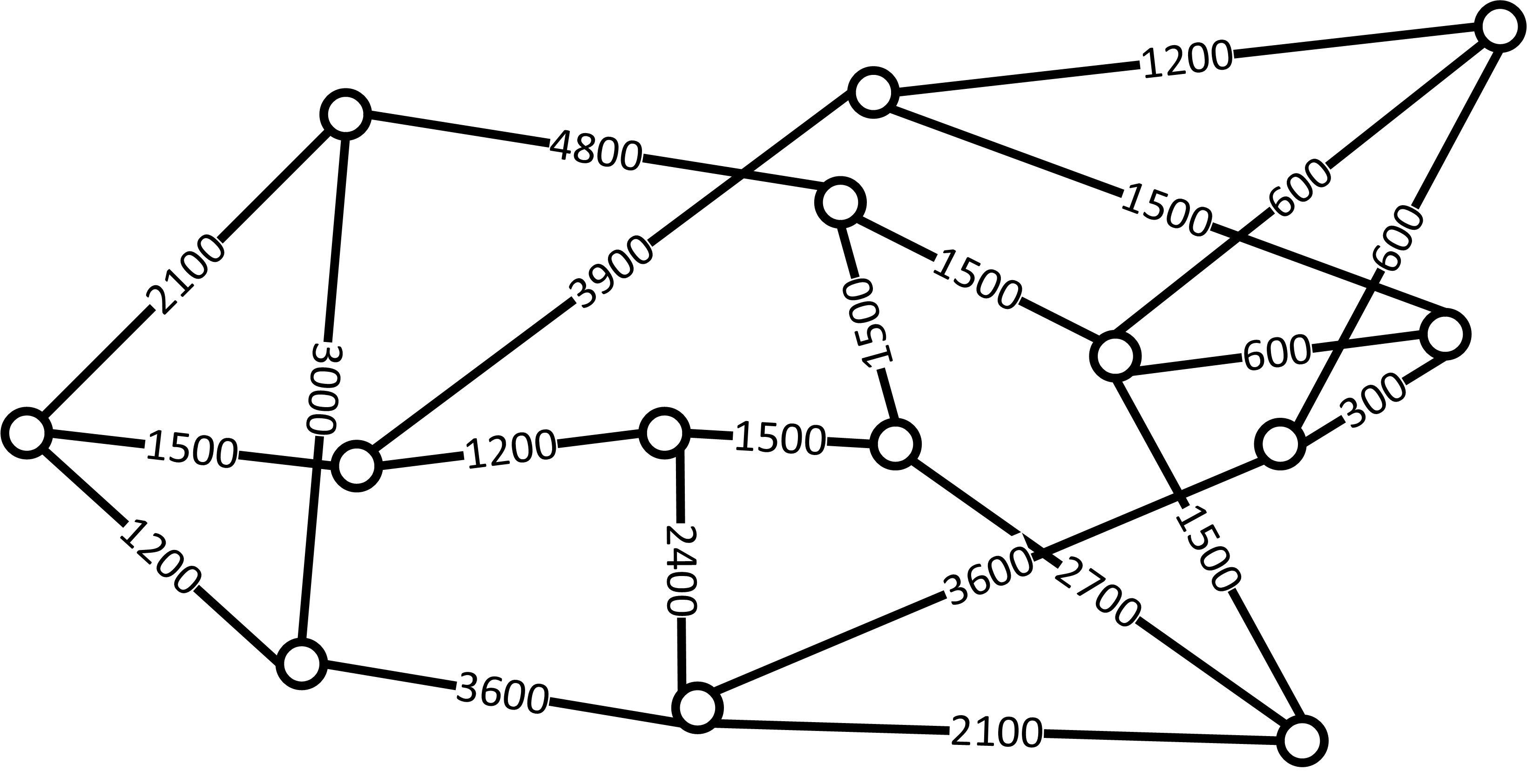}
    \caption{14 nodes, 22 links NSFNET, with distance in km marked on the each edge.}
    \label{fig:e}
\end{figure}
\subsection{Network Settings}
To evaluate the efficacy of our proposed algorithms, we operated a set of simulations experiments using MATLAB R2019b. The simulations are done for 1,00,000 requests with multiple iterations. The performance of proposed algorithms - RSACS with Dynamic Shared Backup Path Protection and Dynamic Cycles in Flexigrid Optical Networks, are evaluated on the 14-nodes 22-links NSFNET having an average degree of all the vertices\footnote{a.k.a. nodal degree, $n_d = \dfrac{2E}{V}$} is 3.0 as shown in figure \ref{fig:e}. We assume the fiber bandwidth to be 4 THz on each link of the network. Using O-OFDM technology, the whole bandwidth is divided into 12.5 GHz parallel channels. Therefore, there are 320 spectrum slots on each link of the network. The traffic demands for all the lightpath requests on each node pair are uniformly distributed. The bandwidth required for each lightpath is chosen randomly in range from 1 to \textit{B} Gbps, where different `\textit{B}' values are used as parameter in simulations. In this thesis, the value of B is 100 Gbps\footnote{8 spectrum slots, if the grid size is 12.5 GHz}. For spectrum allocation, we also considered an additional guard band (GB). The size of GB is considered to be 10 GHz. 

The use of static traffic for simulation does not show the effectiveness of the proposed algorithms. Hence, lightpath requests were  generated dynamically, i.e., we considered dynamic traffic scenario. The incoming lightpaths can be set up and released upon request. These are equivalent to setting up and releasing circuits in circuit-switched networks. The incoming lightpath requests arrive with an exponentially distributed inter-arrival time with the average of $\dfrac{1}{\lambda}$ seconds. Each connection is maintained for exponentially distributed holding time with average of $\dfrac{1}{\mu}$ seconds before being released. The offered load ($\rho$) in Erlang (E) is given by

\begin{equation}
\rho = \dfrac{\dfrac{1}{\mu}}{\dfrac{1}{\lambda}} = \dfrac{\lambda}{\mu}.
\end{equation}

The performance parameters are estimated based on the observations made during the steady-state condition (which is observed to happen after approximately three times the average holding time, i.e., $3 \times\dfrac{1}{\mu}$).

In this, we have used average values of failure rate ($\lambda'$) and repair rate ($\mu'$), such that availability values on all links are uniformly distributed. We are keeping fix the value of failure rate and changing the values of repair rate. The link's availability and threshold availability ranges from 0.9 (one nine) to 0.999999 (six nines) as shown in Table \ref{Table:Ava}.

\begin{table}[ht]
\centering
\small
\begin{tabular}{|c|c|c|c|}
\hline
 \multicolumn{2}{|c|}{Availability Values}\\
\cline{1-2}
one nine & 0.9 \\
\hline
two nines & 0.99\\
\hline
three nines & 0.999\\
\hline
four nines & 0.9999\\
\hline
five nines & 0.99999\\
\hline
six nines & 0.999999\\
\hline
\end{tabular}
 
\caption{Different Availability values.}
\label{Table:Ava}
\end{table} 

\subsection{Simulation Results}
The Type-II algorithm is used for finding the working paths. When the path's availability is lower than the threshold availability, then for backup paths, DSBPSS and D-cycles based on the Type-II algorithm are used. 

\subsubsection{DSBPSS}

\begin{figure*}
	\centering
     \captionsetup{justification=centering}
\begin{subfigure}{0.3\textwidth}
    \centering
    \includegraphics[width=\linewidth]{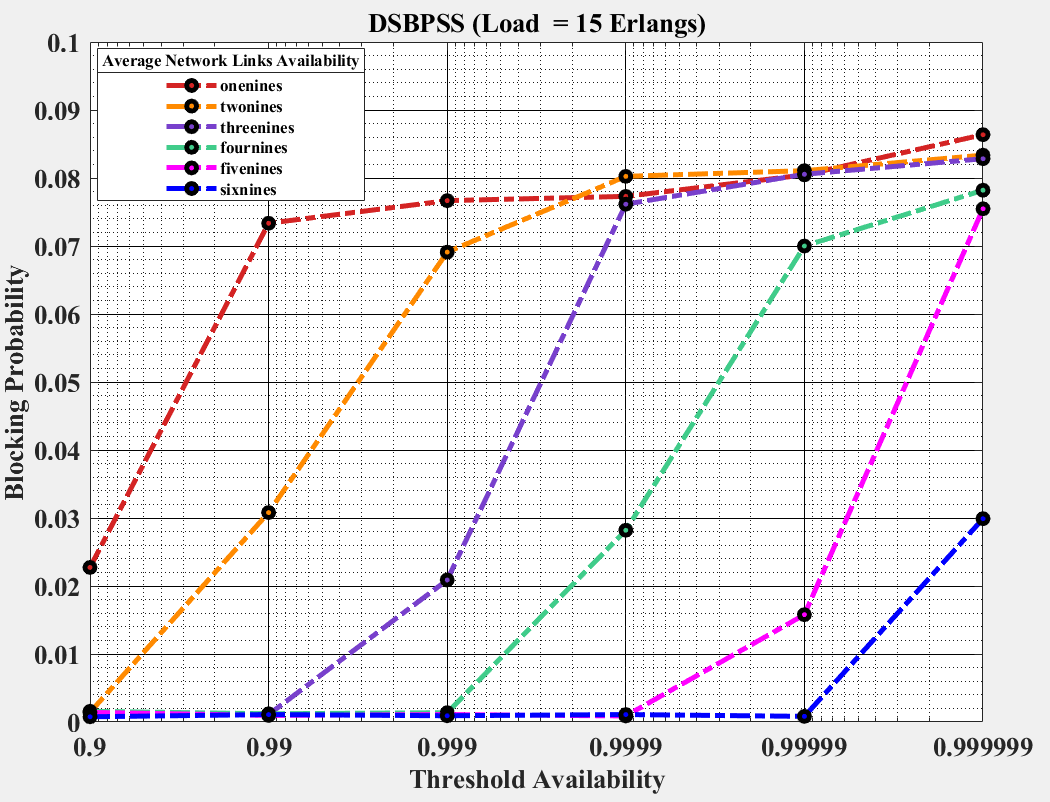}
    \caption{}
    \label{fig:b15s}
\end{subfigure}
\begin{subfigure}{0.3\textwidth}
    \centering
    \includegraphics[width=\linewidth]{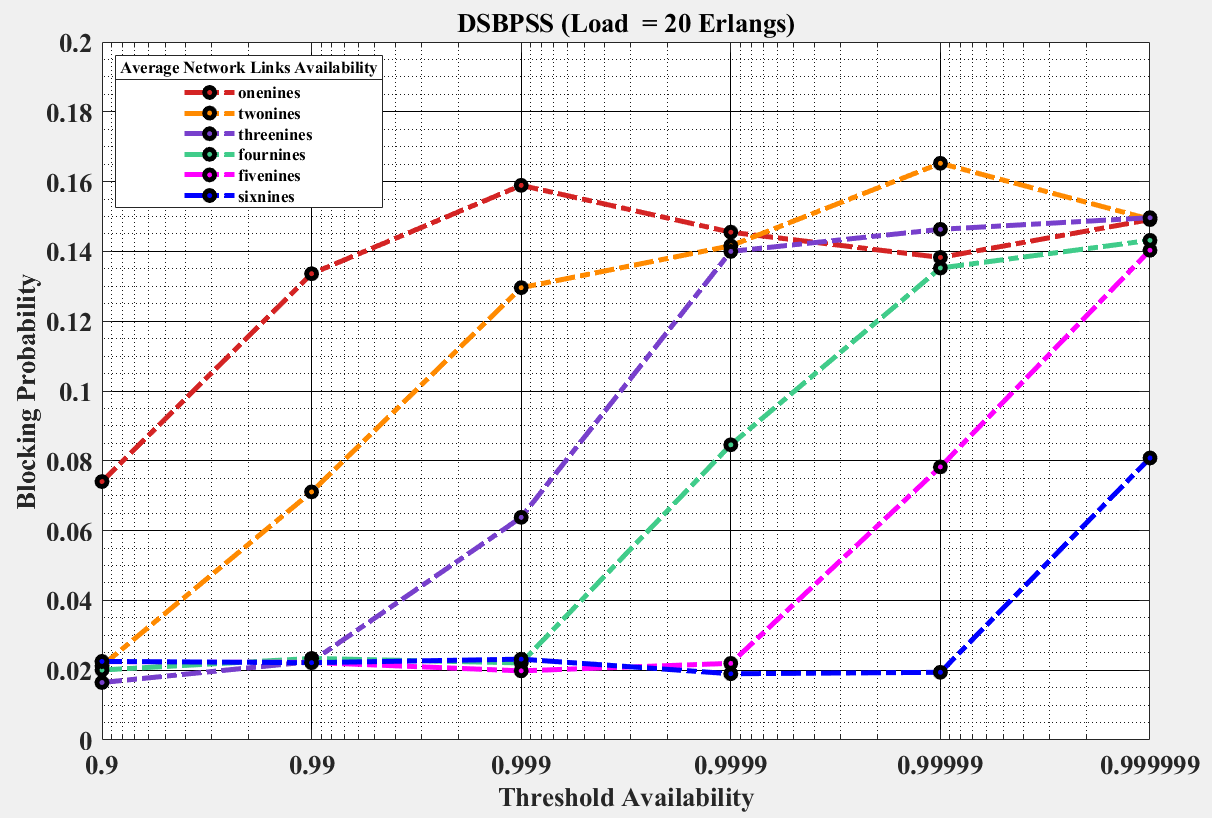}
    \caption{}
    \label{fig:b20s}
\end{subfigure}
\begin{subfigure}{0.3\textwidth}
    \centering
    \includegraphics[width=\linewidth]{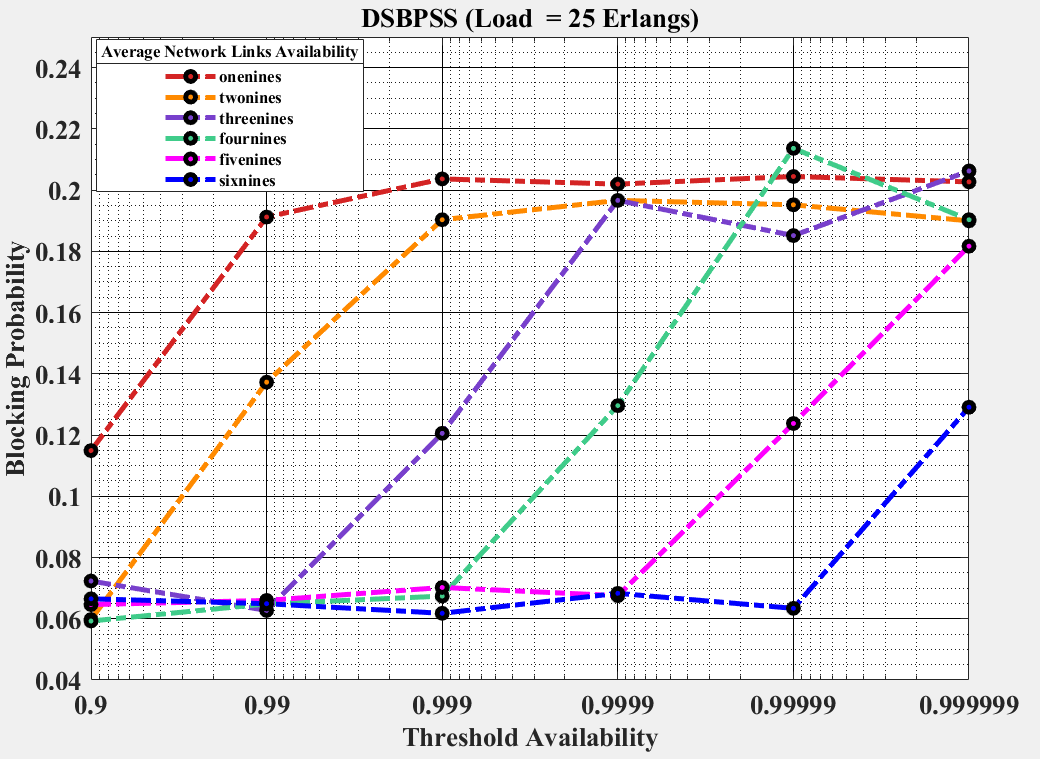}
    \caption{}
    \label{fig:b25s}
\end{subfigure}
\caption{Blocking Probability against various average link availabilities and offered load per node for Type II-DSBPSS method.}
\label{fig:bs}
\end{figure*}

Figure \ref{fig:bs} plots the blocking probability for DSBPPS for different average network links availability and load values. As seen in the figures \ref{fig:b15s}, \ref{fig:b20s}, and \ref{fig:b25s}, the blocking probability is due to the blocking of the working lightpath requests when the Average Network Links Availability $> A_{th}$. Whereas, as the Average Network Links Availability $\leq A_{th}$, the impact of the backup path on blocking probability is also observed in the figures, i.e., the increase in the blocking probability is due to the incoming working path and in ability to setup backup path. Based on the observations, we can infer that when the Average Network Links Availability $\leq A_{th}$, the backup path occupies the available spectrum slots at that instant such that it becomes unavailable for new lightpath requests. 

As the difference between Average Network Links Availability and $A_{th}$ increases, the blocking of the connection requests also increases. After a certain $A_{th}$, the blocking probability reaches a steady-state condition. There is a minor difference between blocking probabilities.

A similar pattern can be observed for figures \ref{fig:b20s}, and \ref{fig:b25s}, as the offered load per node changes. The contrast is as the load increases, the blocking probability also increases for various values of Average Network Links Availability and $A_{th}$. 
We can also observe, after a certain threshold value for offered load per node of 20 Erlang blocking starts decreasing. Most likely, it is happening because backup paths could not be setup, leaving free spaces for more connections. At $A_{th}$ = 0.999, more backup path could be setup. This phenomena will be indent at higher load.

\begin{figure*}
\centering
     \captionsetup{justification=centering}
\begin{subfigure}{0.3\textwidth}
    \centering
    \includegraphics[width=\linewidth]{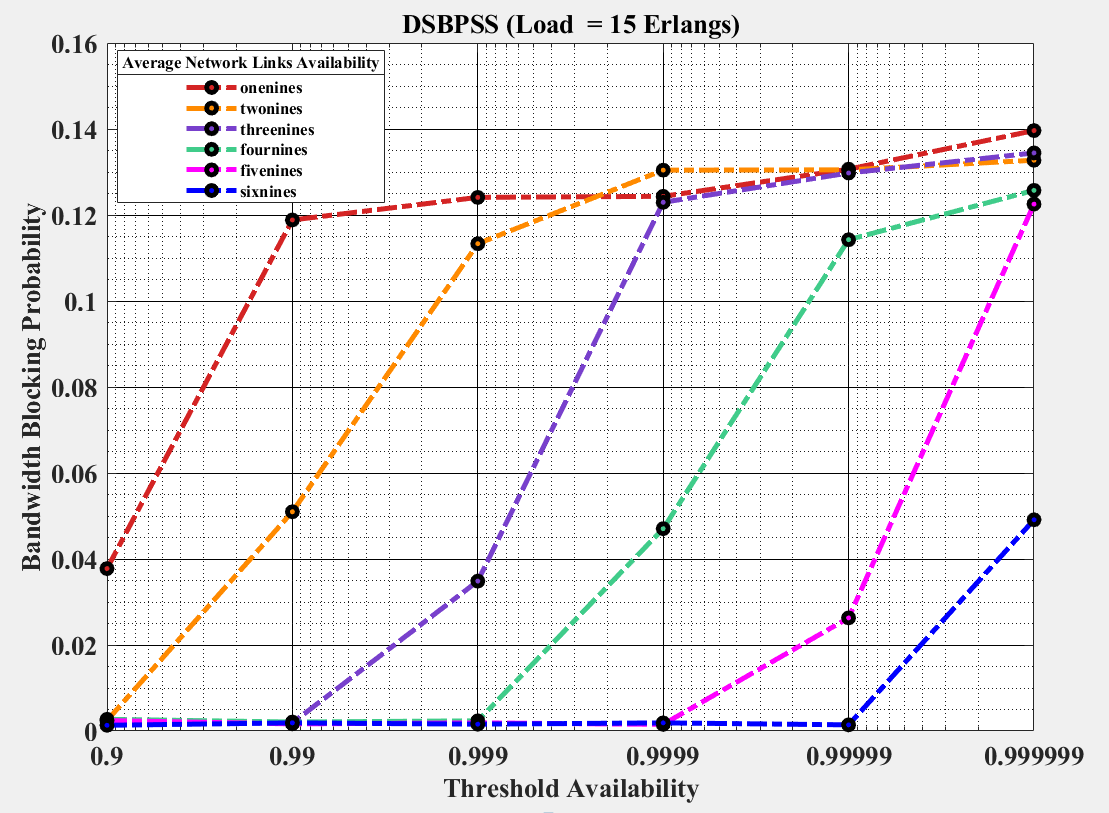}
    \caption{}
    \label{fig:bbp15s}
\end{subfigure}
\begin{subfigure}{0.3\textwidth}
    \centering
    \includegraphics[width=\linewidth]{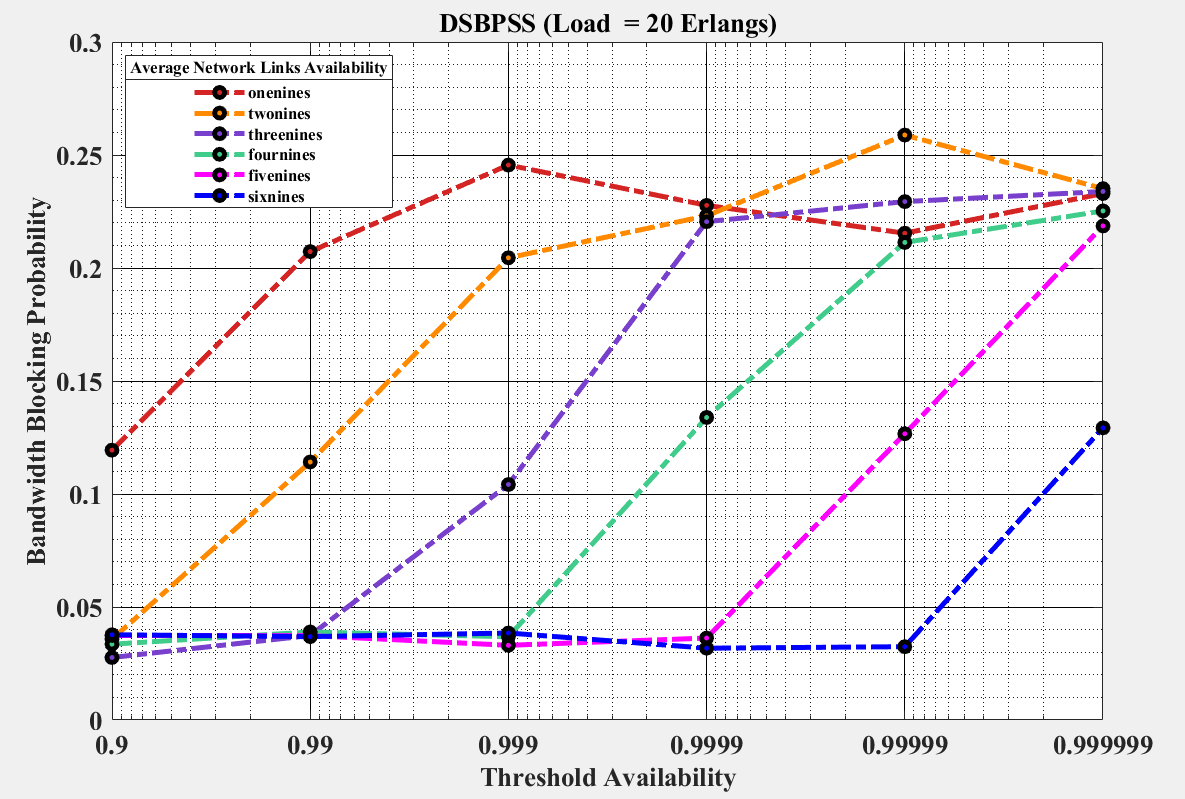}
    \caption{}
    \label{fig:bbp20s}
\end{subfigure}
\begin{subfigure}{0.3\textwidth}
    \centering
    \includegraphics[width=\linewidth]{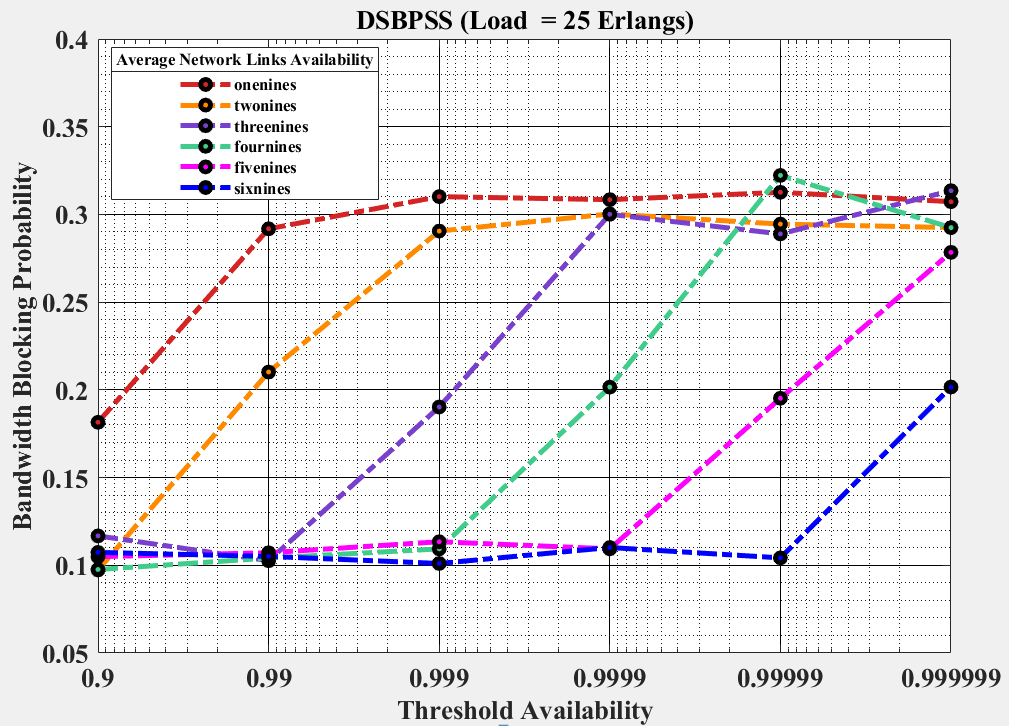}
    \caption{}
    \label{fig:bbp25s}
\end{subfigure}
\caption{Bandwidth Blocking Probability against various average link availabilities and offered load per node for Type II-DSBPSS method.}
\label{fig:bbps}
\end{figure*}

\begin{figure*}
\centering
     \captionsetup{justification=centering}
\begin{subfigure}{0.3\textwidth}
    \centering
    \includegraphics[width=\linewidth]{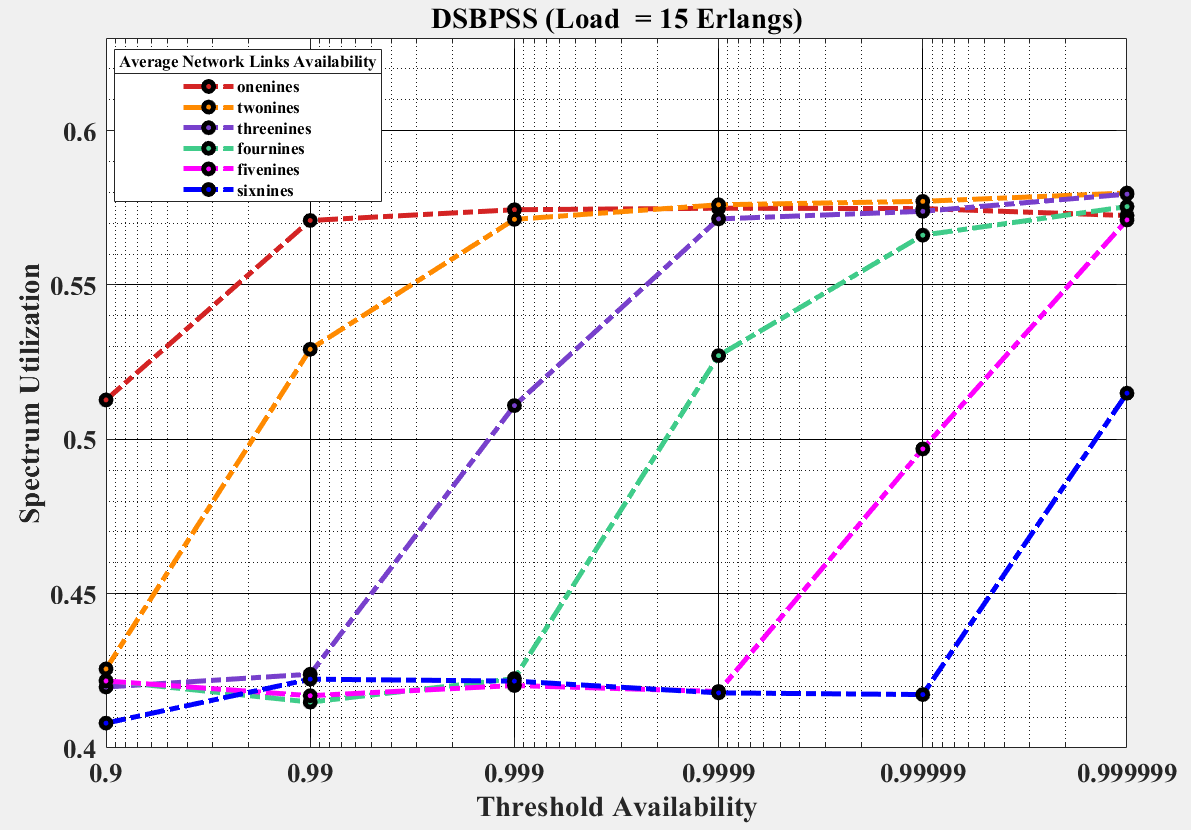}
    \caption{}
    \label{fig:n15s}
\end{subfigure}
\begin{subfigure}{0.3\textwidth}
    \centering
    \includegraphics[width=\linewidth]{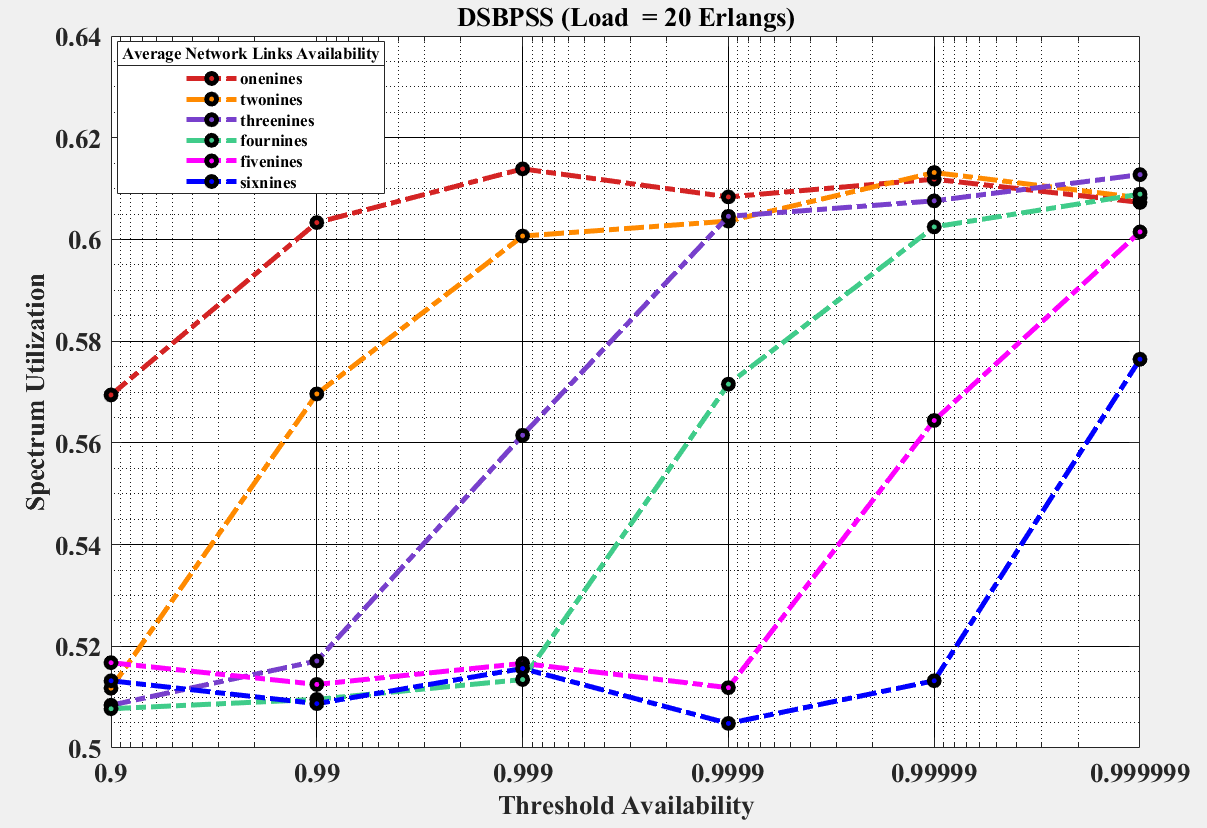}
    \caption{}
    \label{fig:n20s}
\end{subfigure}
\begin{subfigure}{0.3\textwidth}
    \centering
    \includegraphics[width=\linewidth]{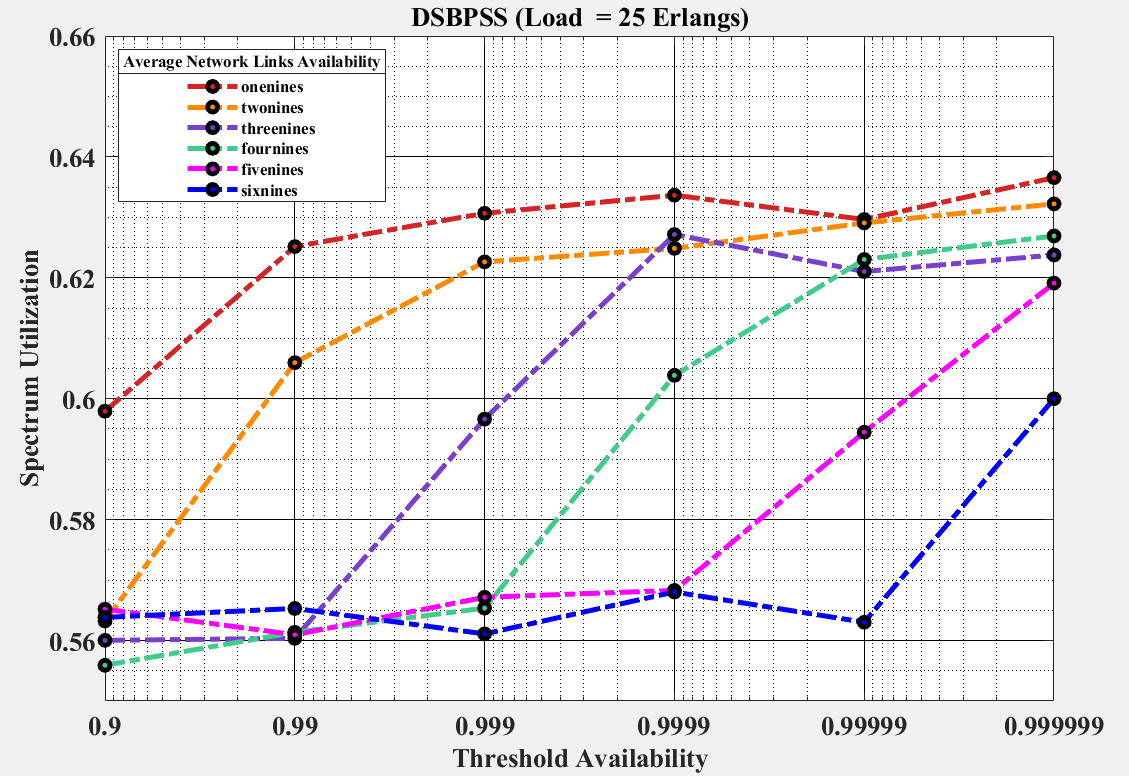}
    \caption{}
    \label{fig:n25s}
\end{subfigure}
\caption{Spectrum Utilization against various average link availabilities and offered load per node for Type II-DSBPSS method.}
\label{fig:ns}
\end{figure*}

\begin{figure*}
\centering
     \captionsetup{justification=centering}
\begin{subfigure}{0.3\textwidth}
    \centering
    \includegraphics[width=\linewidth]{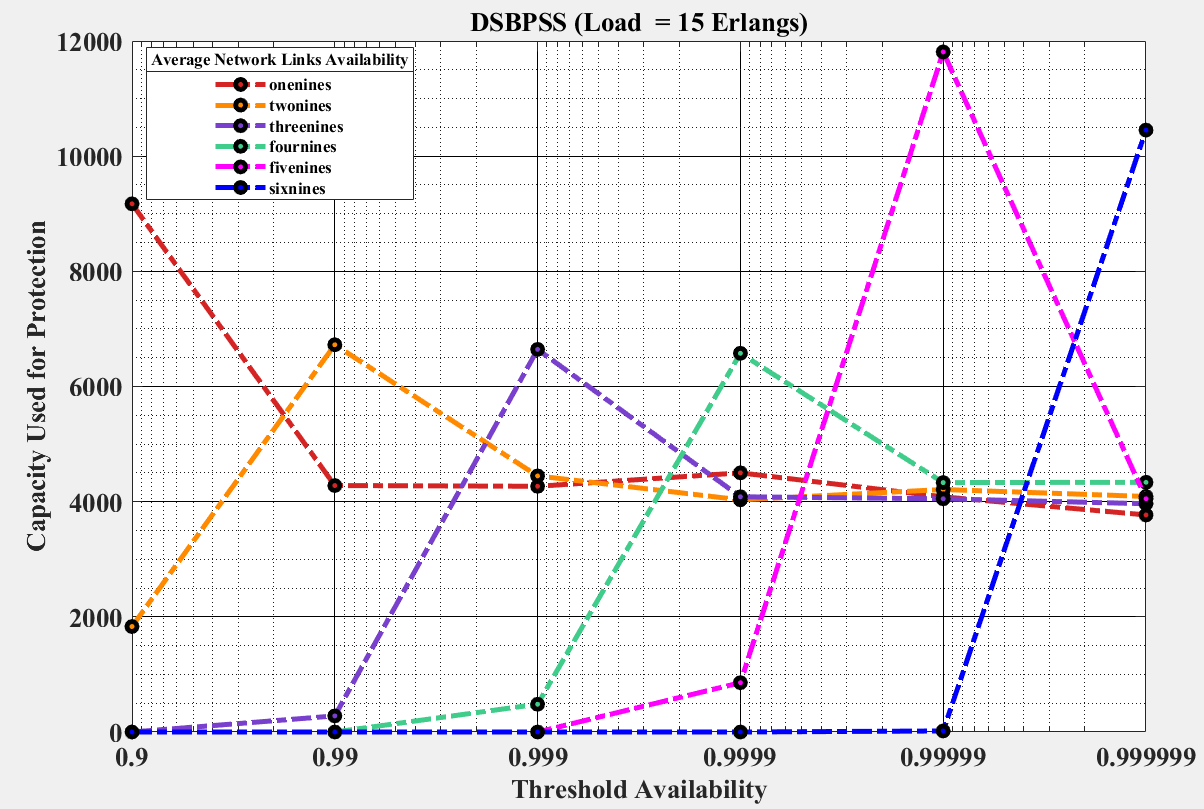}
    \caption{}
    \label{fig:sc15s}
\end{subfigure}
\begin{subfigure}{0.3\textwidth}
    \centering
    \includegraphics[width=\linewidth]{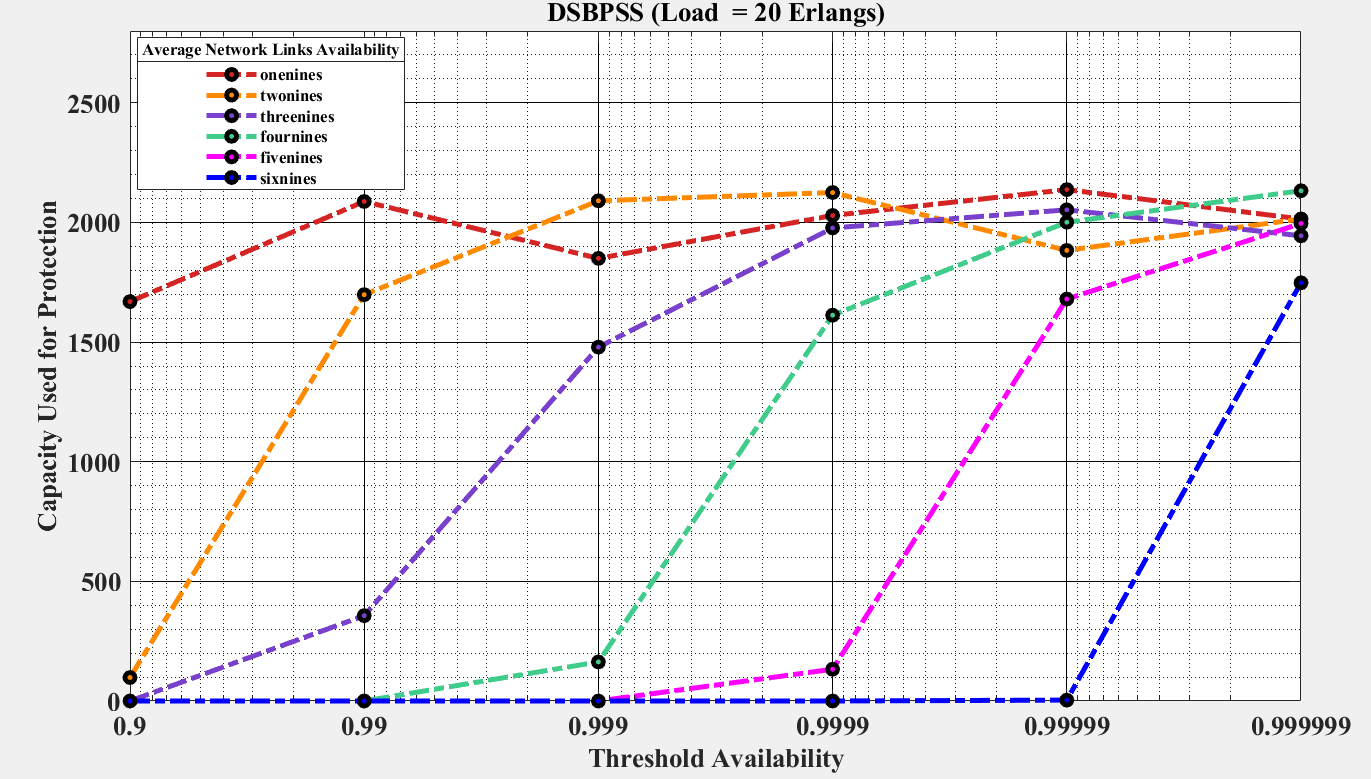}
    \caption{}
    \label{fig:sc20s}
\end{subfigure}
\begin{subfigure}{0.3\textwidth}
    \centering
    \includegraphics[width=\linewidth]{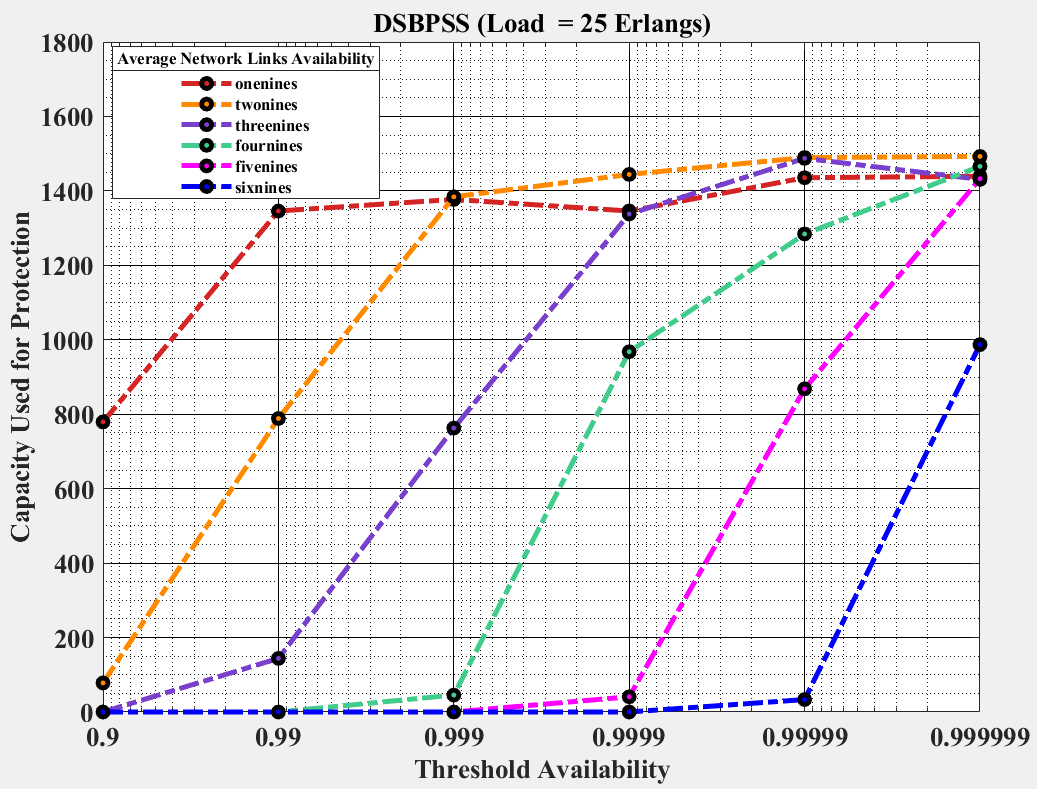}
    \caption{}
    \label{fig:sc25s}
\end{subfigure}
\caption{Capacity Used for Protection against various average link availabilities and offered load per node for Type II-DSBPSS method.}
\label{fig:scs}
\end{figure*}

\begin{figure*}
\centering
     \captionsetup{justification=centering}
\begin{subfigure}{0.3\textwidth}
    \centering
    \includegraphics[width=\linewidth]{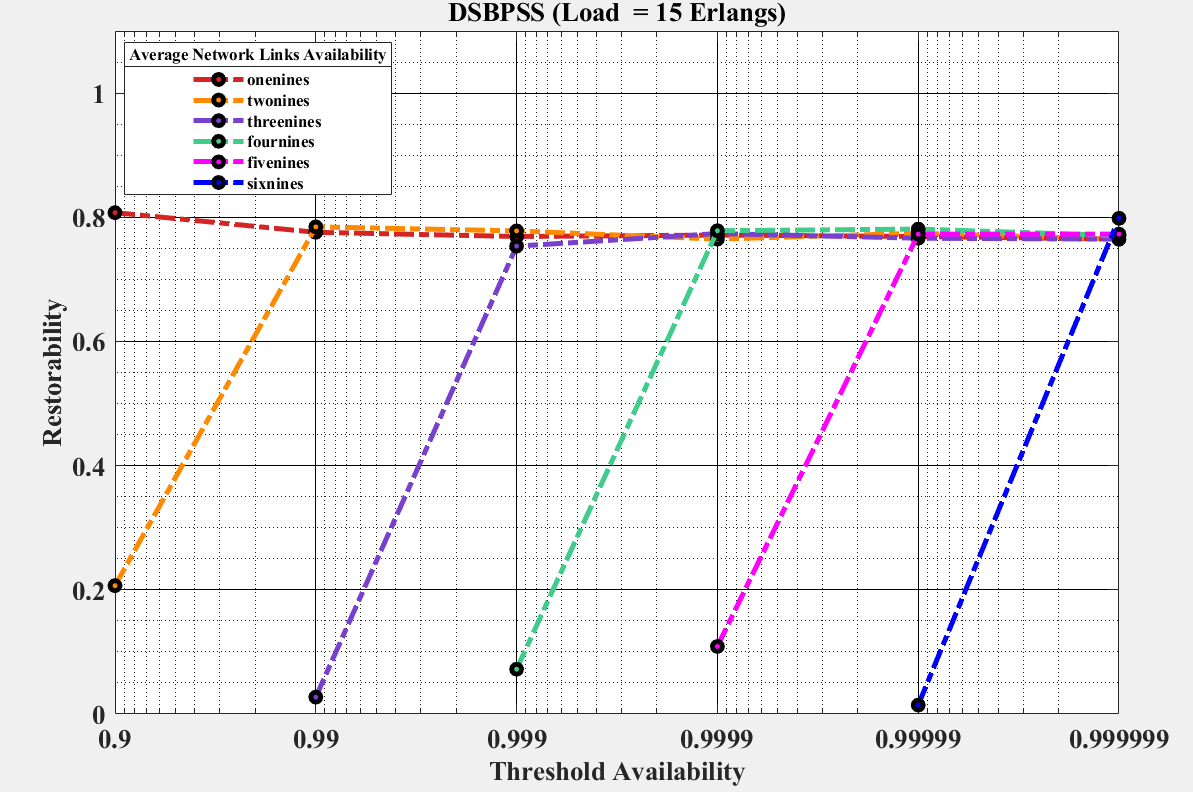}
    \caption{}
    \label{fig:r15s}
\end{subfigure}
\begin{subfigure}{0.3\textwidth}
    \centering
    \includegraphics[width=\linewidth]{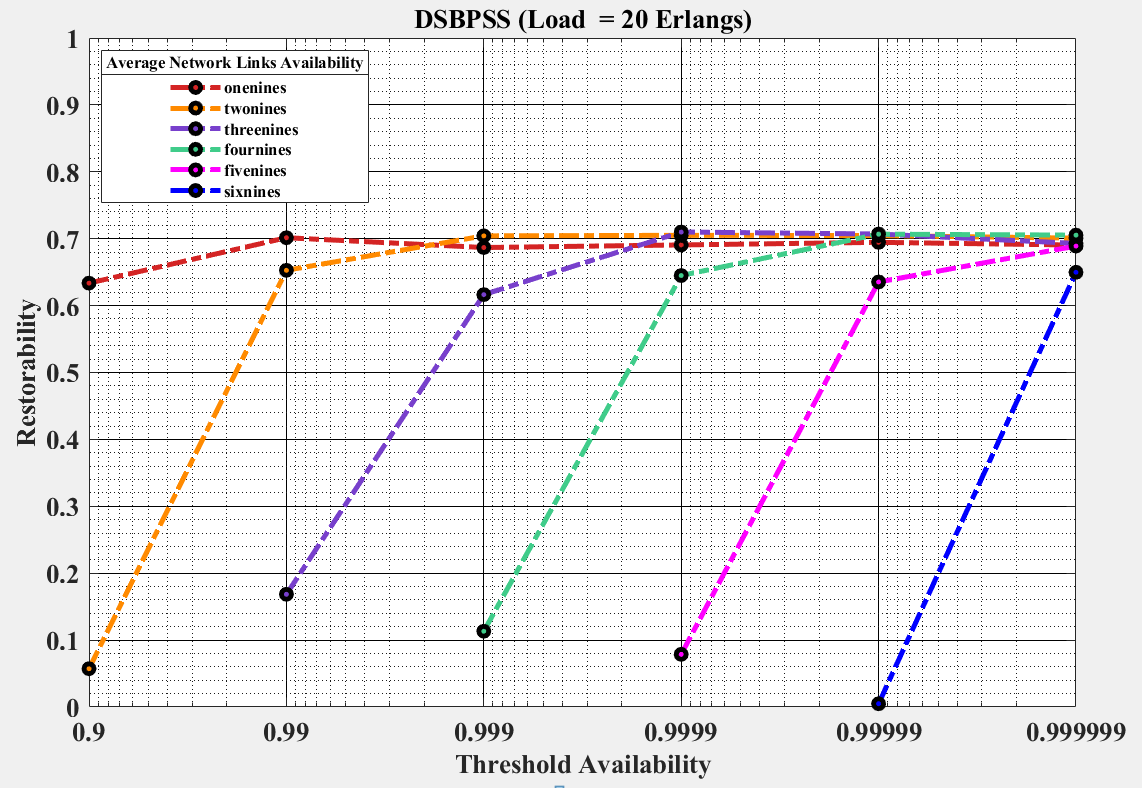}
    \caption{}
    \label{fig:r20s}
\end{subfigure}
\begin{subfigure}{0.3\textwidth}
    \centering
    \includegraphics[width=\linewidth]{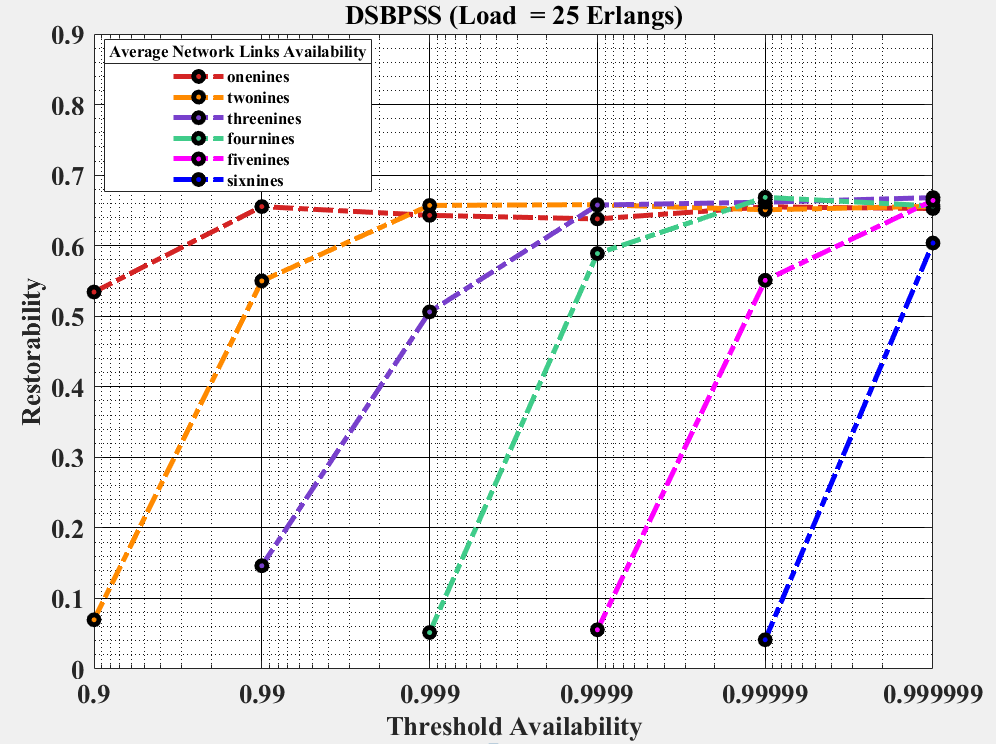}
    \caption{}
    \label{fig:r25s}
\end{subfigure}
\caption{Restorability against various average link availabilities and offered load per node for Type II-DSBPSS method.}
\label{fig:rs}
\end{figure*}
Figure \ref{fig:bbps} is the bandwidth blocking probability for DSBPPS for different average network links availability and load values. The conclusions are the same for blocking probability, except the bandwidth blocking probability is slightly higher for all the slot sizes. It happens because instead of counting the number of blocked connections, we consider each lightpath request's bandwidth. 

Figure \ref{fig:ns} is the spectrum utilization for DSBPPS for different average network links availability and load values. As the Average Network Links Availability $ \leq Ath $, the impact of the backup paths on the spectrum utilization is observed in the figures \ref{fig:ns}. The difference between Average Network Links Availability and $A_{th}$ progresses, the utilization of the spectrum increases due to backup path and working path. After a certain $A_{th}$, the spectrum utilization reaches a steady-state condition. 

The plots in Figure \ref{fig:scs} are for capacity used for protection by backup path for DSBPPS against different average network links availability and load values. When the Average Network Links Availability $ > Ath $, zero capacity is required for protection. Now, as the Average Network Links Availability $ \leq Ath $, backup paths are required, which increases capacity used for protection by the backup paths. The redundant paths are provisioned to the working path till the $A_{pp}^{max} \geq A_{th} $. As the difference between Average Network Links Availability and $A_{th}$ progresses, the capacity required for protection starts decreasing due to the unavailability of the spectrum slots. After a certain $A_{th}$, it reaches a steady-state condition. As the load values increases, the steady-state condition is maintained as achieved earlier. 

The plots in Figure \ref{fig:rs} are for restorability of lightpath requests whose $A_{pp}^{max} < A_{th}$ using DSBPPS. These are plotted against different average network links availability and load values. When the Average Network Links Availability $ > A_{th} $, no protection is required; therefore, from here, we can interpret that there is no restorability ratio. Now, as the Average Network Links Availability $ \leq A_{th} $ backup paths are required, which results in requirement of more protection paths. Since restorability is the ratio of the total number of protected paths to the total number of paths requiring protection (sum of the protected and unprotected paths), this ratio is only for the paths whose Availability is less than the threshold Availability.  Now, as the requirement for protection increases, the number of the protected paths also increases. Hence the restorability ratio also increases. The redundant paths are provisioned to the working path till the $A_{pp}^{max} \geq A_{th} $. As the difference between Average Network Links Availability and $A_{th}$ progresses, the restorability ratio decreases due to the unavailability of the spectrum slots. Therefore, the denominator term (protected paths $+$ unprotected paths) starts dominating. This means there is an increase in the number of unprotected paths. After a certain $A_{th}$, it reaches a steady-state condition. As the load values increases further, the steady-state condition is maintained as achieved earlier. 

The increase in the protection of number of lightpath requests can be done by increasing the capacity of each link in the network, multiple core-based networks or multiple band optical transmission networks. 

\subsubsection{D-Cycles}

\begin{figure*}
\centering
     \captionsetup{justification=centering}
\begin{subfigure}{0.3\textwidth}
    \centering
    \includegraphics[width=\linewidth]{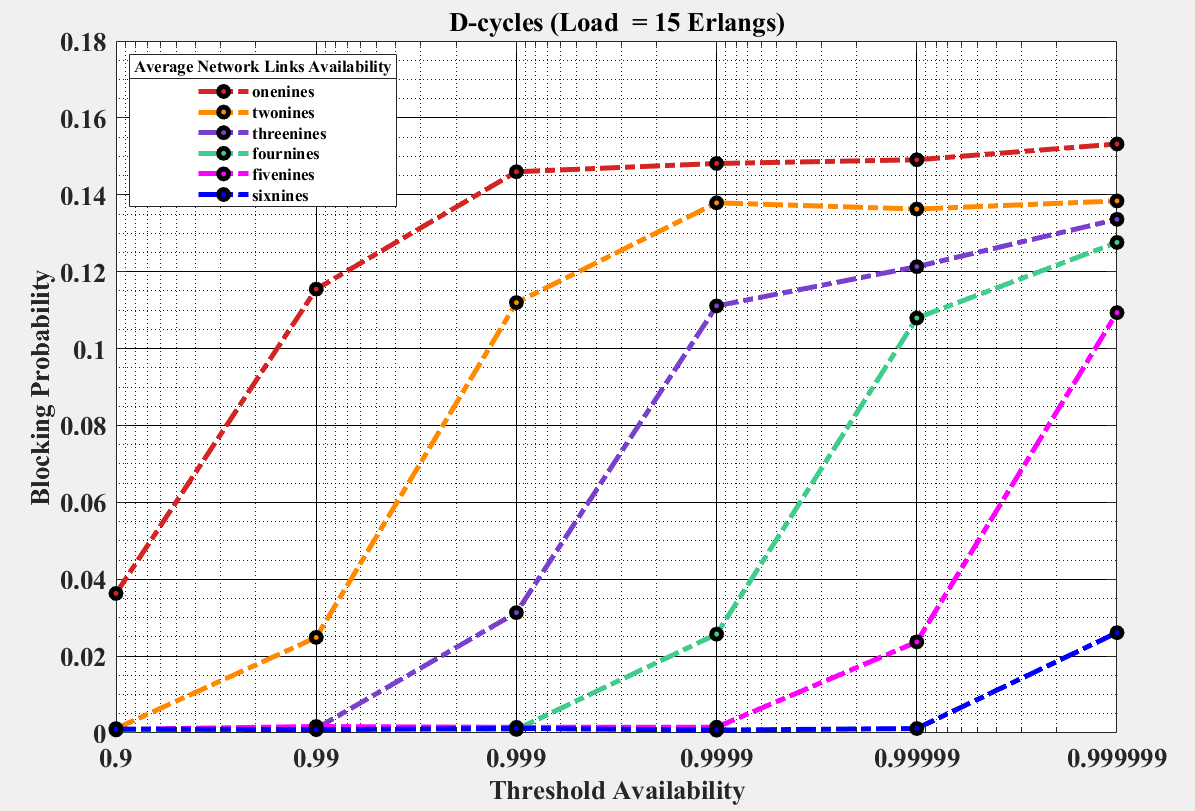}
    \caption{}
    \label{fig:b15d}
\end{subfigure}
\begin{subfigure}{0.3\textwidth}
    \centering
    \includegraphics[width=\linewidth]{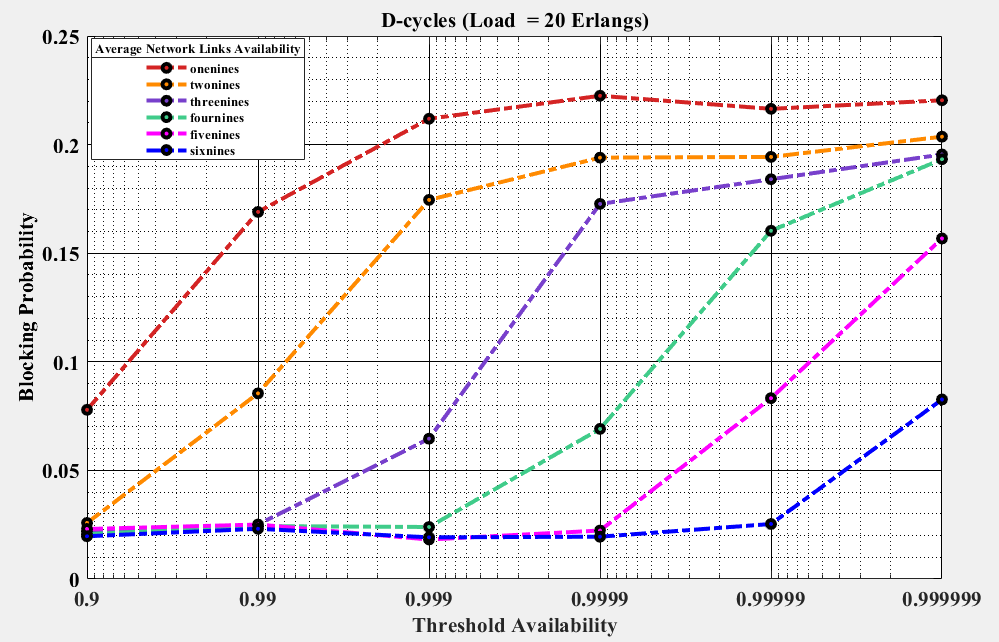}
    \caption{}
    \label{fig:b20d}
\end{subfigure}
\begin{subfigure}{0.3\textwidth}
    \centering
    \includegraphics[width=\linewidth]{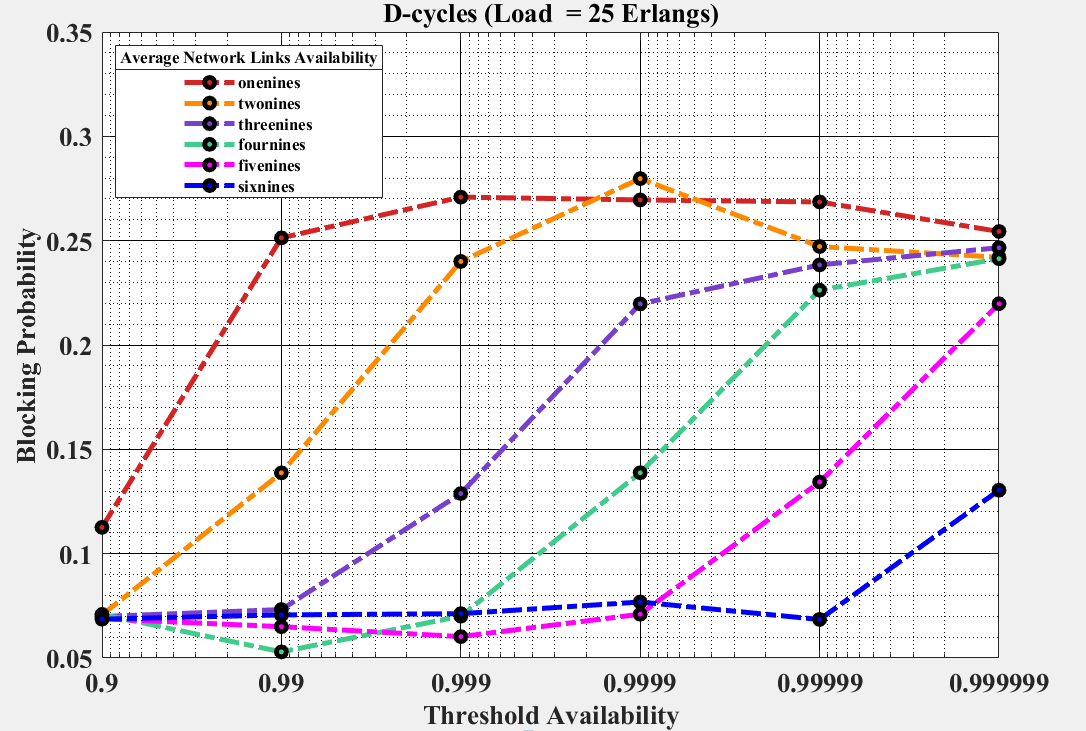}
    \caption{}
    \label{fig:b25d}
\end{subfigure}
\caption{Blocking Probability against various average link availabilities and offered load per node for Type II-D-cycles method.}
\label{fig:bd}
\end{figure*}

Figure \ref{fig:bd} plots the blocking probability for D-cycles for different average network links availability and load values. As seen in the figures \ref{fig:b15d}, \ref{fig:b20d}, and \ref{fig:b25d}, the blocking probability is due to the blocking of the working lightpath requests when the Average Network Links Availability $> A_{th}$. Whereas, as the Average Network Links Availability $\leq A_{th}$, the impact of the protection on blocking probability is also observed in the figures, i.e., the increase in the blocking probability is due to the incoming working path and backup D-cycles. Based on the observations, when the Average Network Links Availability $\leq A_{th}$, the part of the D-cycles used for protection occupies the available spectrum slots at that instant such that it becomes unavailable for new lightpath requests.

As the difference between Average Network Links Availability and $A_{th}$ increases, the blocking of the connection requests also increases. After a certain $A_{th}$, the blocking probability reaches a steady-state condition. There are minor difference between blocking probabilities.

A similar pattern can be observed for figures \ref{fig:b20d}, and \ref{fig:b25d}, as the offered load per node changes. The contrast is, as the load increases, the blocking probability also increases for various values of Average Network Links Availability and $A_{th}$.

\begin{figure*}
\centering
     \captionsetup{justification=centering}
\begin{subfigure}{0.3\textwidth}
    \centering
    \includegraphics[width=\linewidth]{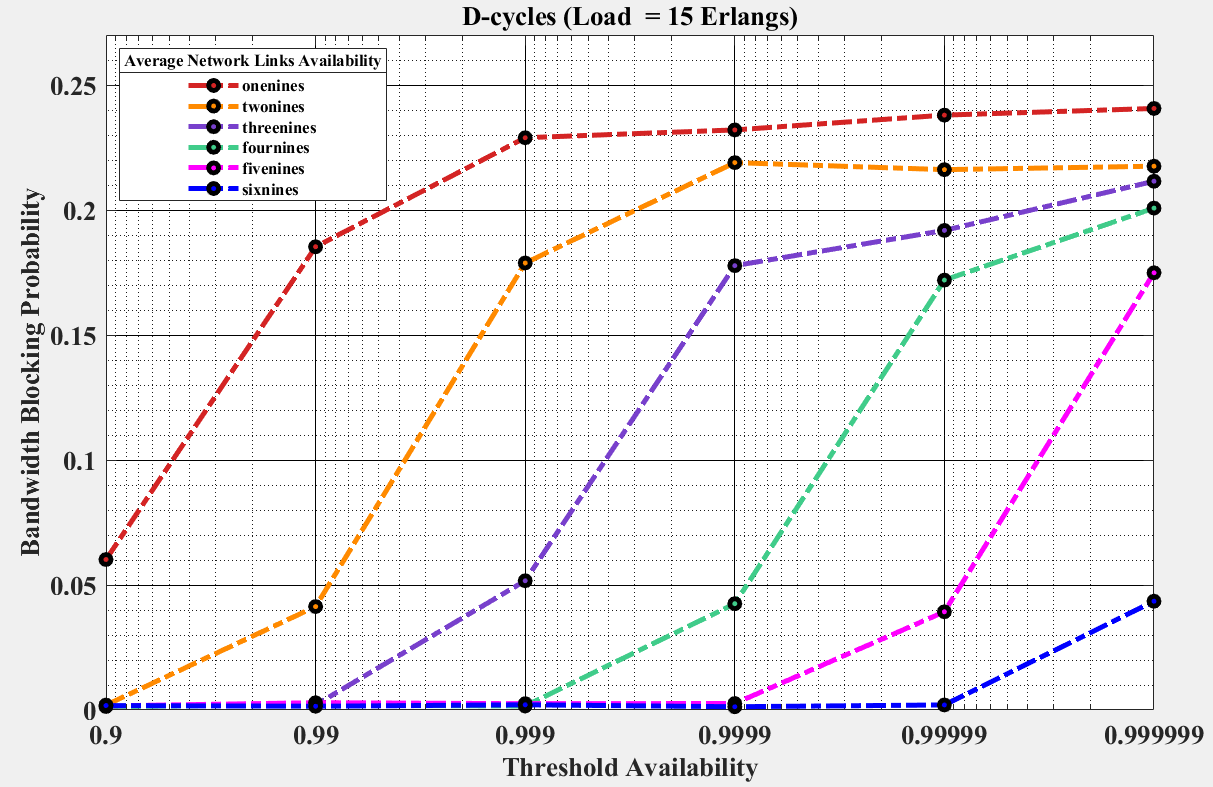}
    \caption{}
    \label{fig:bbp15d}
\end{subfigure}
\begin{subfigure}{0.3\textwidth}
    \centering
    \includegraphics[width=\linewidth]{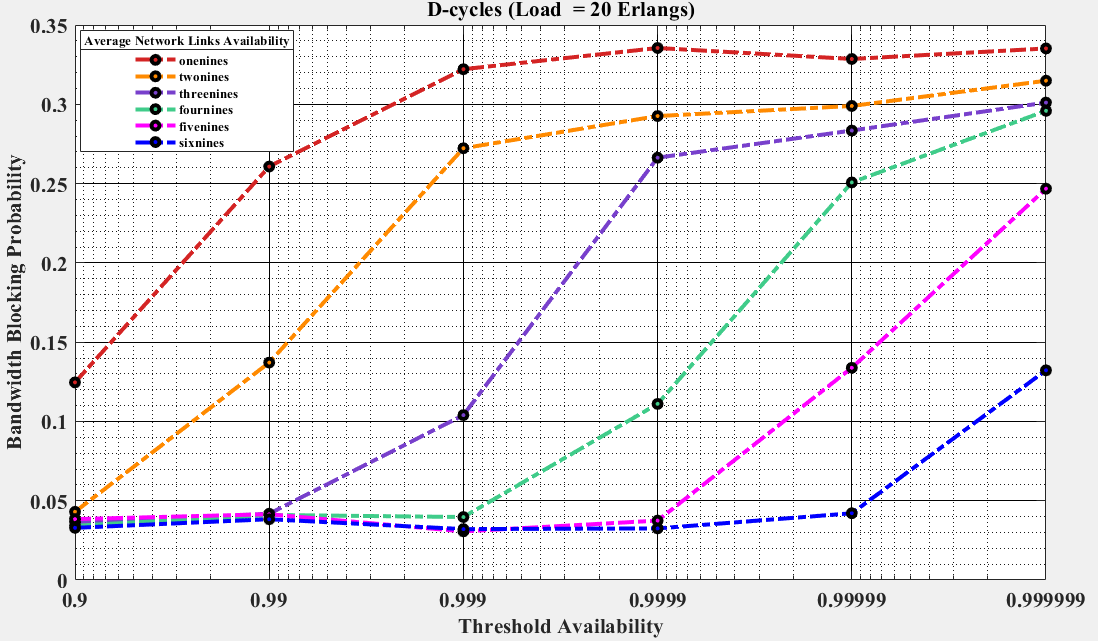}
    \caption{}
    \label{fig:bbp20d}
\end{subfigure}
\begin{subfigure}{0.3\textwidth}
    \centering
    \includegraphics[width=\linewidth]{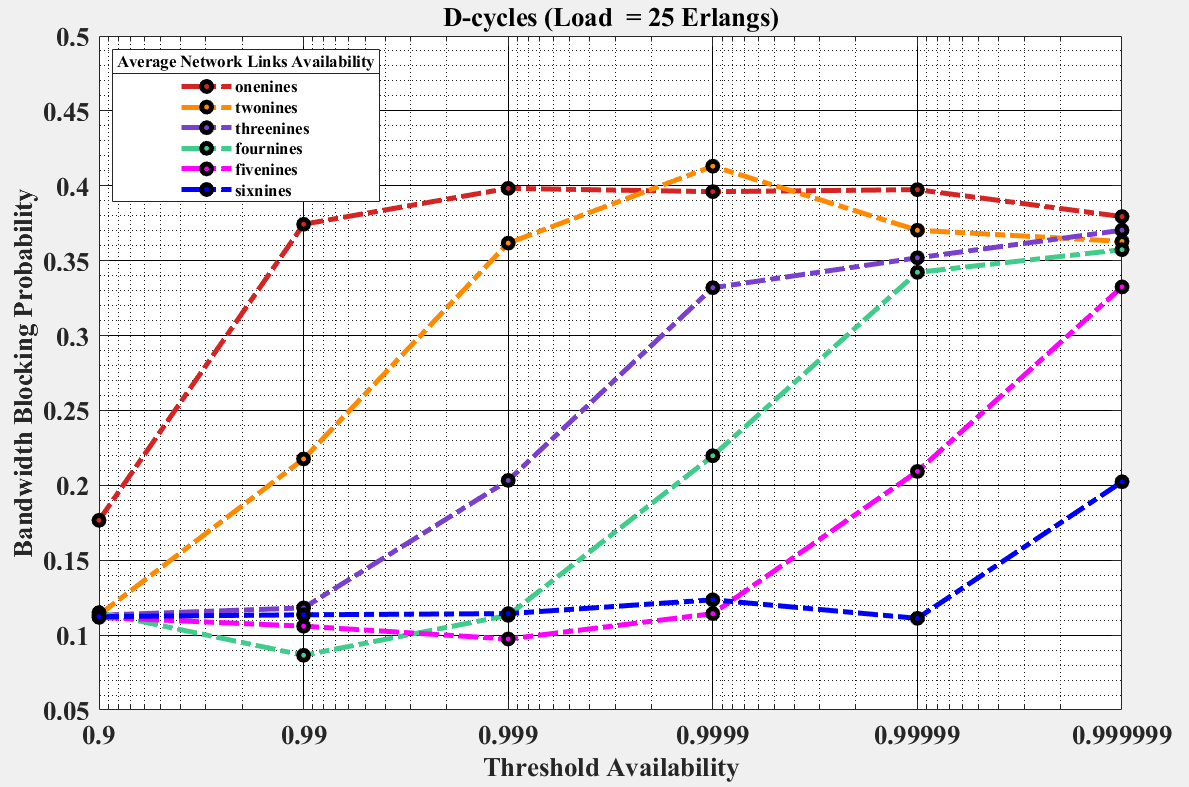}
    \caption{}
    \label{fig:bbp25d}
\end{subfigure}
\caption{Bandwidth Blocking Probability against various average link availabilities and offered load per node for Type II-D-cycles method.}
\label{fig:bbpd}
\end{figure*}

\begin{figure*}
\centering
     \captionsetup{justification=centering}
\begin{subfigure}{0.3\textwidth}
    \centering
    \includegraphics[width=\linewidth]{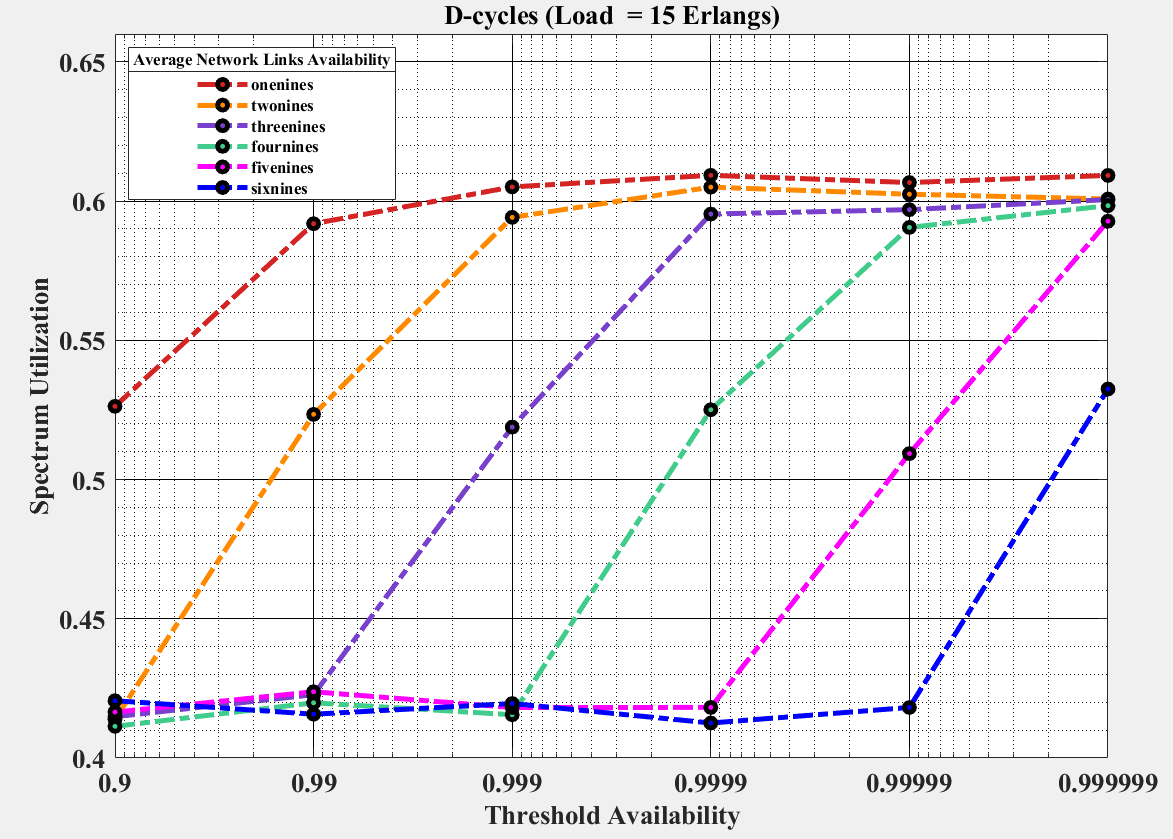}
    \caption{}
    \label{fig:n15d}
\end{subfigure}
\begin{subfigure}{0.3\textwidth}
    \centering
    \includegraphics[width=\linewidth]{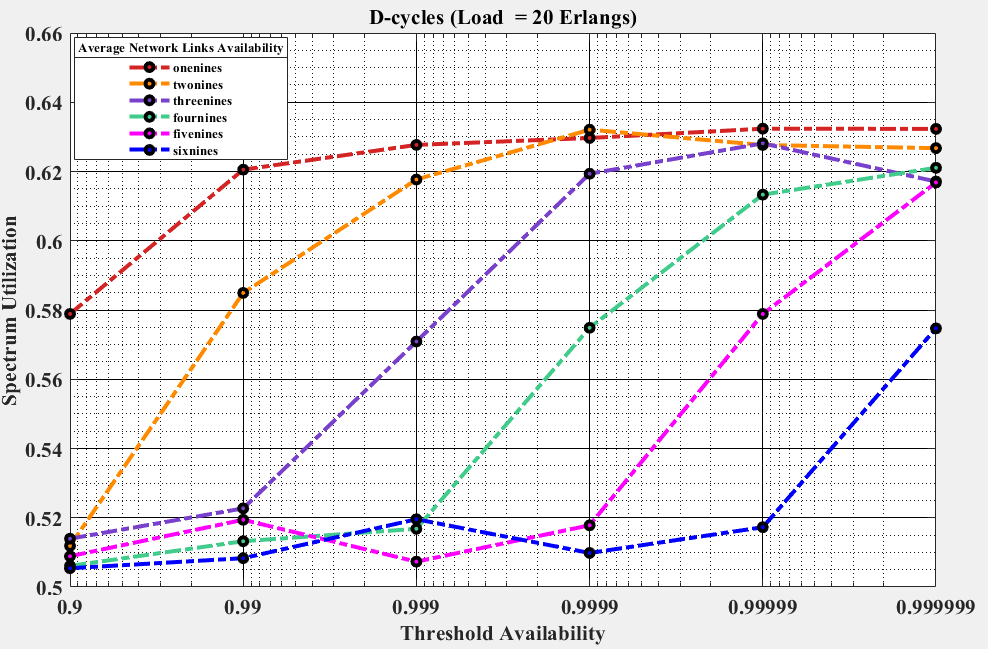}
    \caption{}
    \label{fig:n20d}
\end{subfigure}
\begin{subfigure}{0.3\textwidth}
    \centering
    \includegraphics[width=\linewidth]{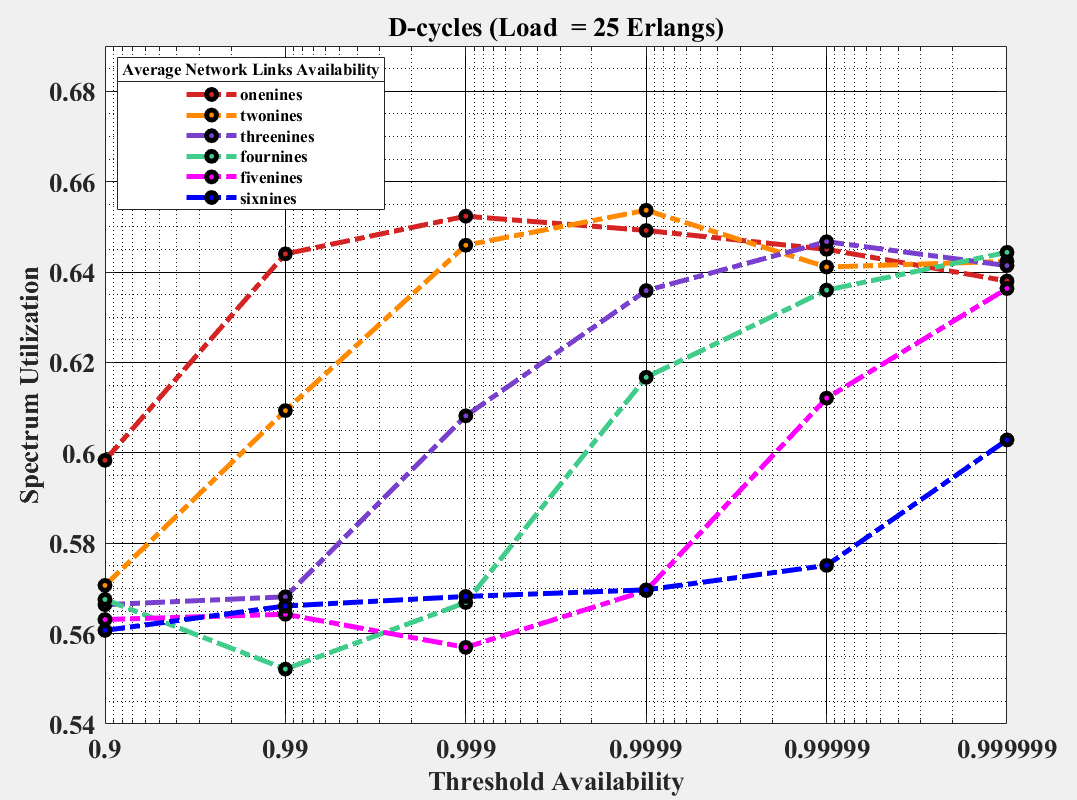}
    \caption{}
    \label{fig:n25d}
\end{subfigure}
\caption{Spectrum Utilization against various average link availabilities and offered load per node for Type II-D-cycles method.}
\label{fig:nd}
\end{figure*}

\begin{figure*}
\centering
     \captionsetup{justification=centering}
\begin{subfigure}{0.3\textwidth}
    \centering
    \includegraphics[width=\linewidth]{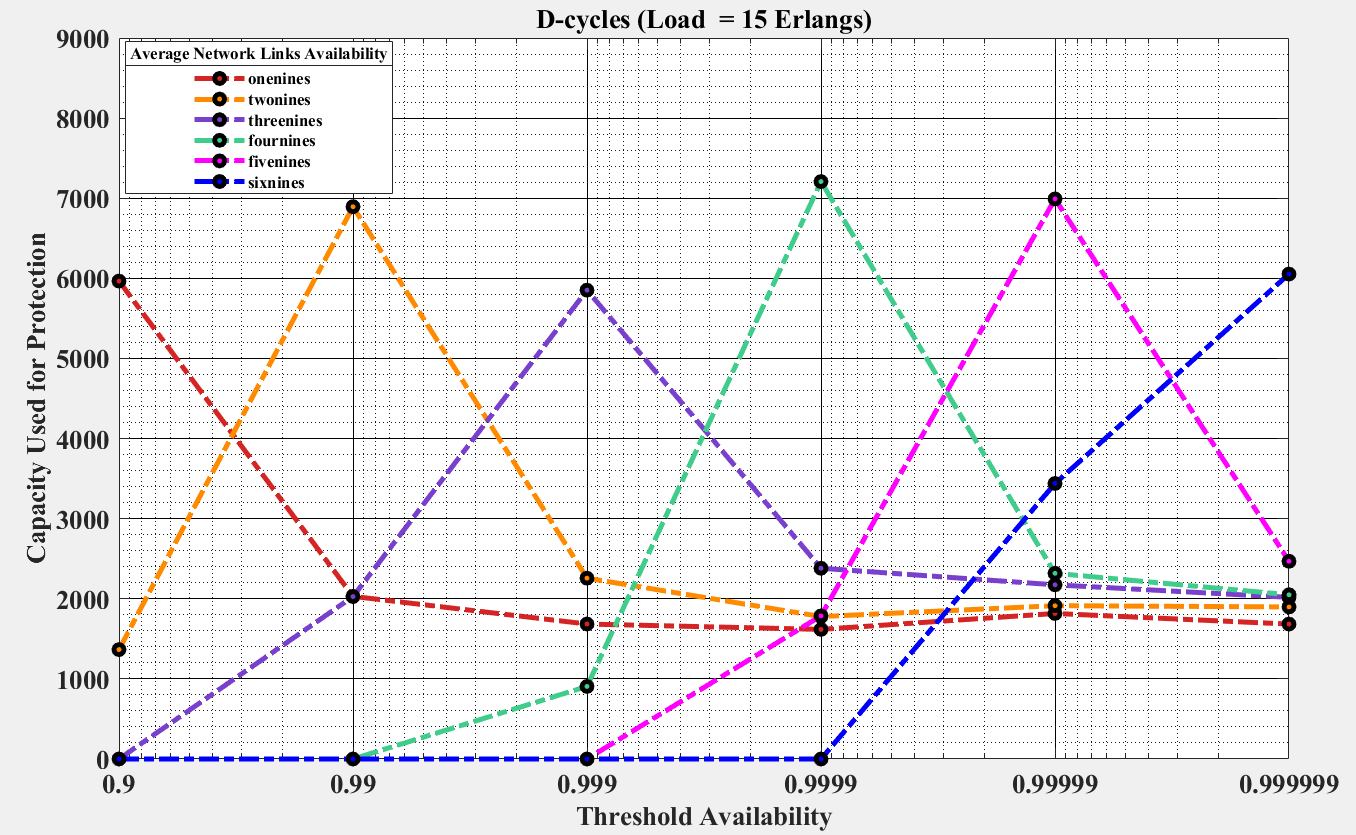}
    \caption{}
    \label{fig:sc15d}
\end{subfigure}
\begin{subfigure}{0.3\textwidth}
    \centering
    \includegraphics[width=\linewidth]{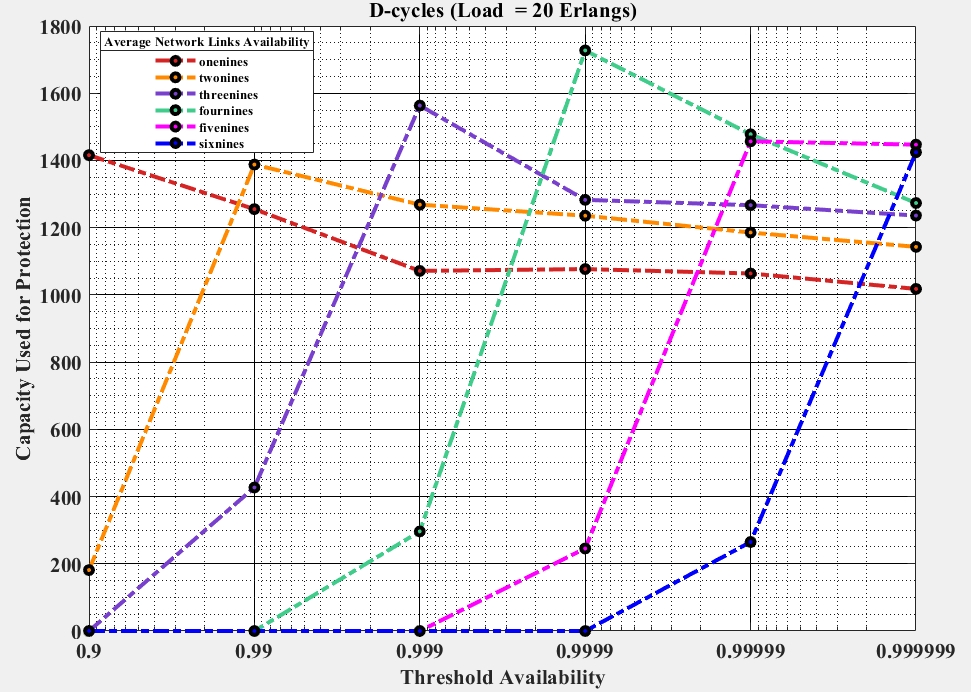}
    \caption{}
    \label{fig:sc20d}
\end{subfigure}
\begin{subfigure}{0.3\textwidth}
    \centering
    \includegraphics[width=\linewidth]{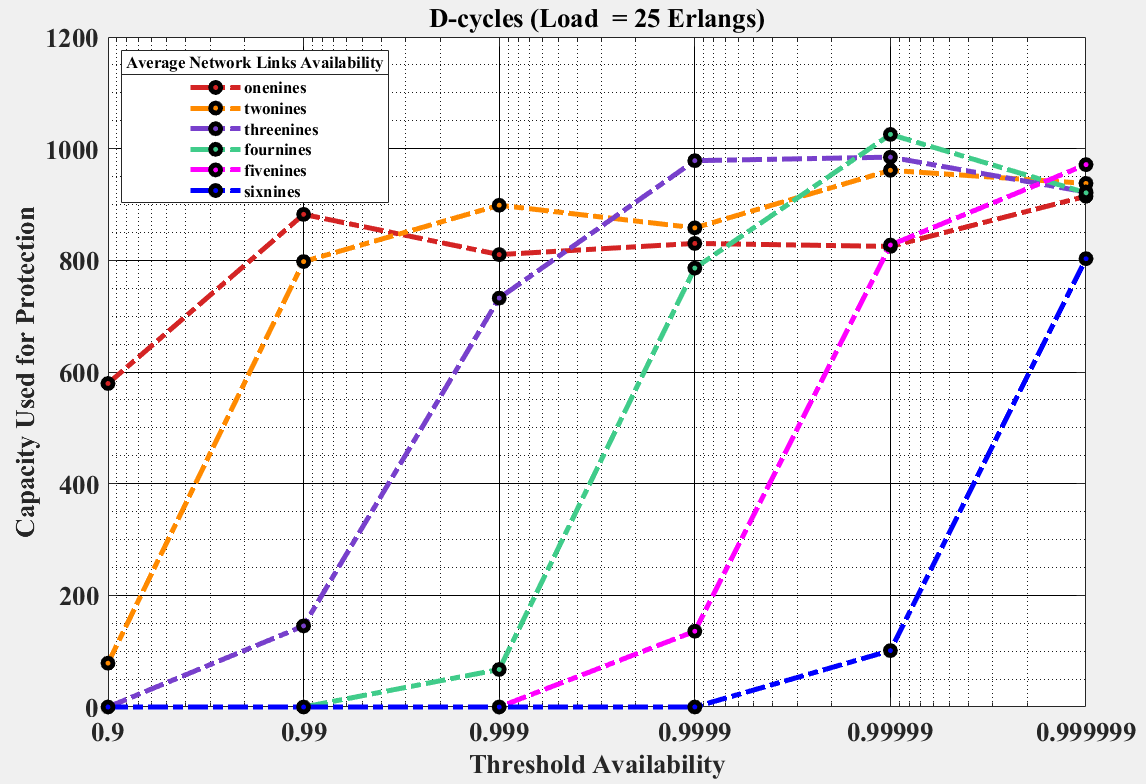}
    \caption{}
    \label{fig:sc25d}
\end{subfigure}
\caption{Capacity Used for Protection against various average link availabilities and offered load per node for Type II-D-cycles method.}
\label{fig:scd}
\end{figure*}

\begin{figure*}
\centering
     \captionsetup{justification=centering}
\begin{subfigure}{0.3\textwidth}
    \centering
    \includegraphics[width=\linewidth]{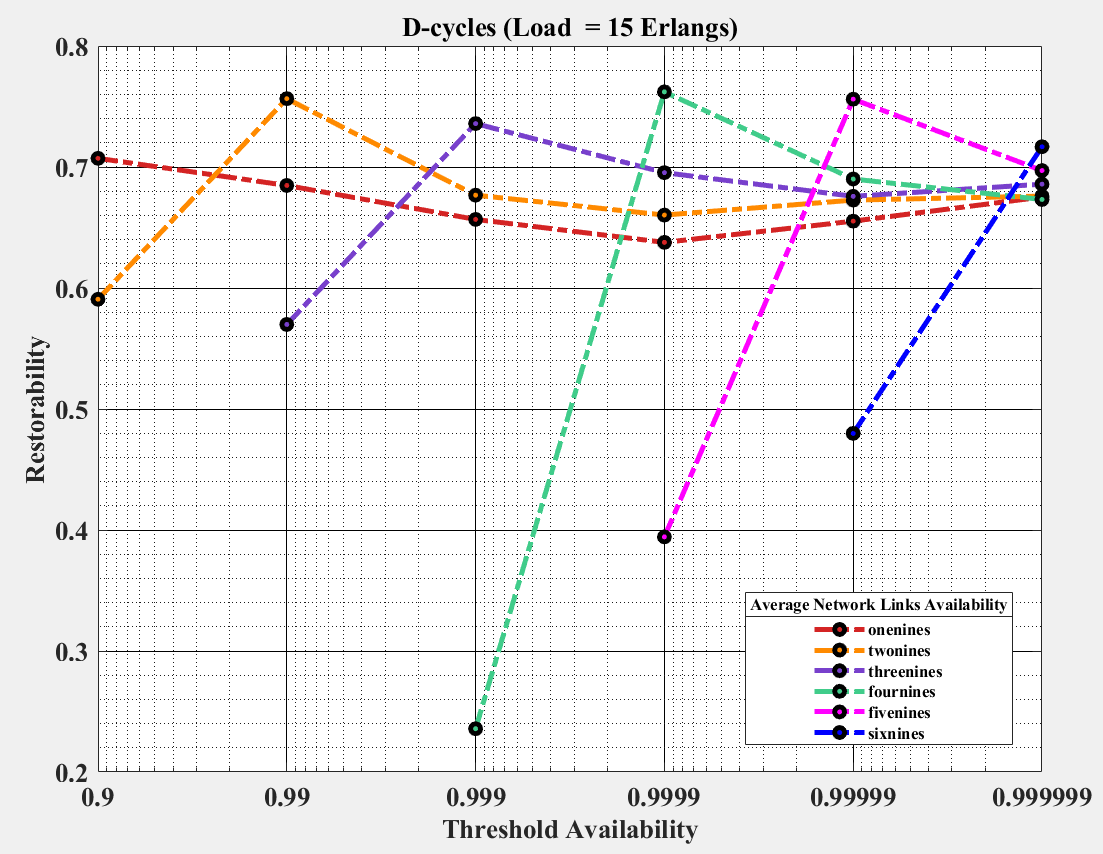}
    \caption{}
    \label{fig:r15d}
\end{subfigure}
\begin{subfigure}{0.3\textwidth}
    \centering
    \includegraphics[width=\linewidth]{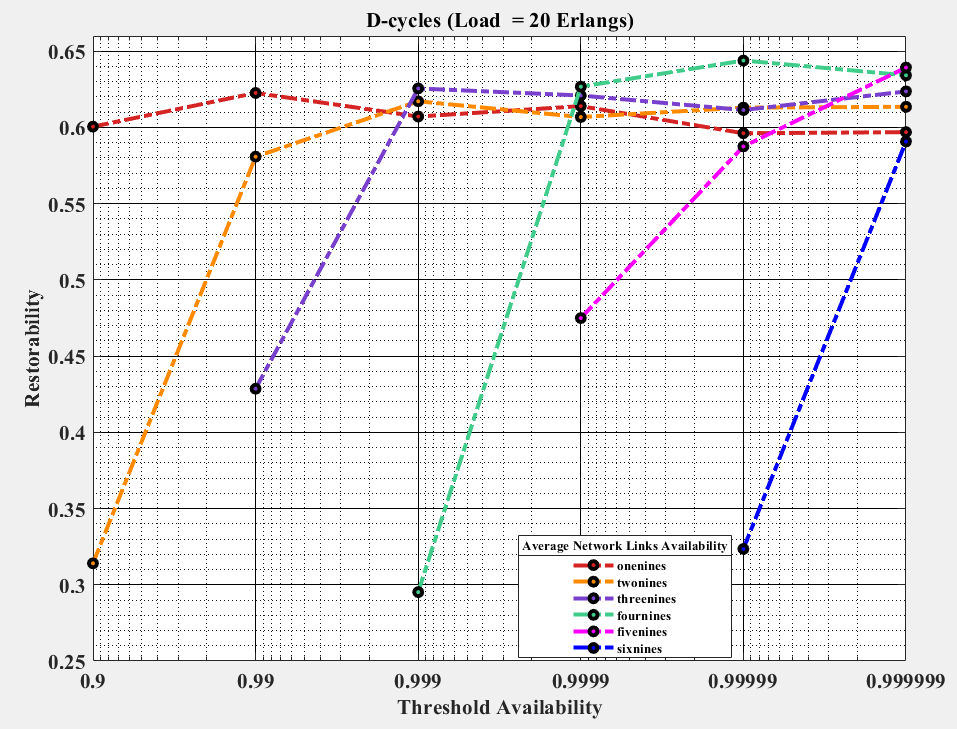}
    \caption{}
    \label{fig:r20d}
\end{subfigure}
\begin{subfigure}{0.3\textwidth}
    \centering
    \includegraphics[width=\linewidth]{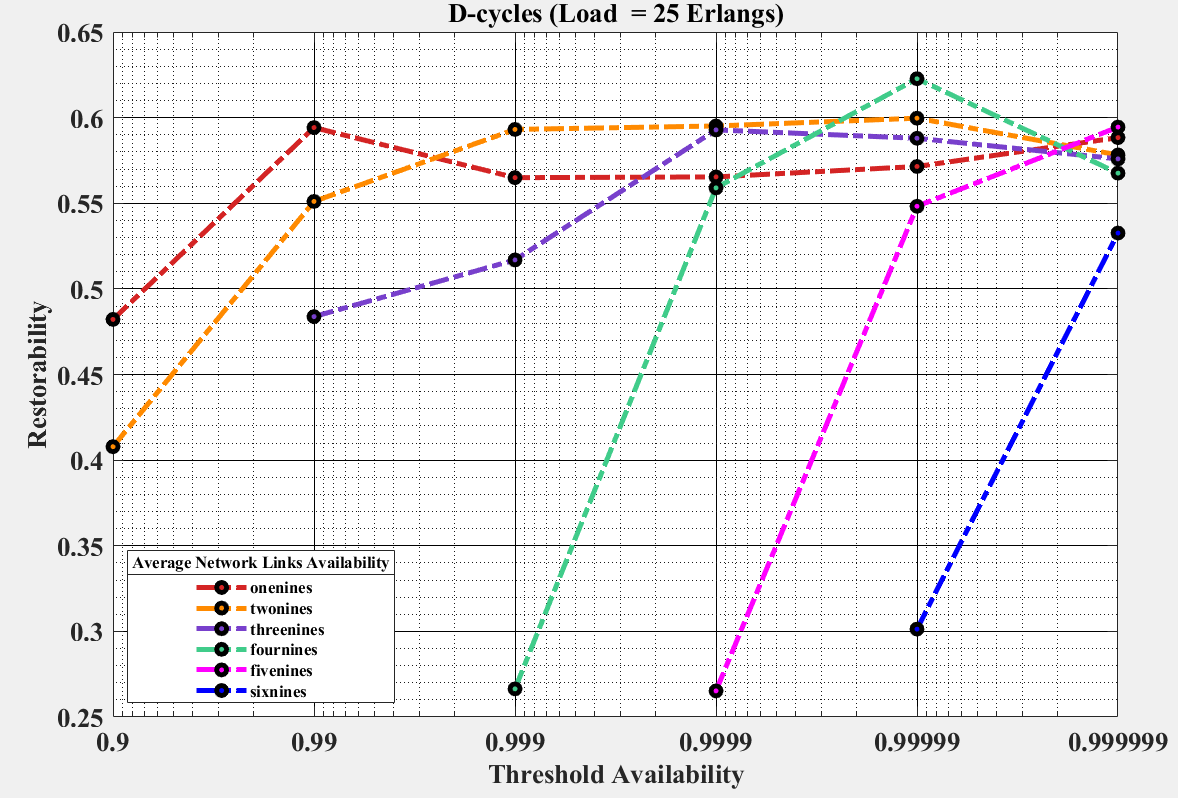}
    \caption{}
    \label{fig:r25d}
\end{subfigure}
\caption{Restorability against various average link availabilities and offered load per node for Type II-D-cycles method.}
\label{fig:rd}
\end{figure*}

Figure \ref{fig:bbpd} is the bandwidth blocking probability for D-cycles for different average network links availability and load values. The conclusions are the same for blocking probability, except the bandwidth blocking probability is slightly higher for all the slot sizes. It happens because instead of counting the number of blocked connections, we consider each lightpath request's bandwidth. 

Figure \ref{fig:nd} is the spectrum utilization for D-cycles for different average network links availability and load values. As the Average Network Links Availability $ \leq Ath $, the impact of the D-cycles requirement on the spectrum utilization is observed in the figures \ref{fig:ns}. The difference between Average Network Links Availability and $A_{th}$ progresses, the utilization of the spectrum due to D-cycles and working path increases. After a certain $A_{th}$, the spectrum utilization reaches a steady-state condition. 

The plots in Figure \ref{fig:scd} are for capacity used for protection using D-cycles. These are plotted against different average network links availability and load values. When the Average Network Links Availability $ > Ath $, zero capacity is required for protection. Now, as the Average Network Links Availability $ \leq Ath $ backup D-cycles are required, which increases capacity used for protection. The redundant paths are provisioned to the link (part of the working path) with minimum Availability till the $A_{pp}^{max} \geq A_{th} $. As the difference between Average Network Links Availability and $A_{th}$ progresses, the capacity required for protection starts decreasing due to the unavailability of the spectrum slots. After a certain $A_{th}$, it reaches a steady-state condition. As the load values increases, the steady-state condition is achieved earlier. 

The plots in Figure \ref{fig:rd} are for restorability of lightpath requests whose $A_{pp}^{max} < A_{th}$ using D-cycles. These are plotted against different average network links availability and load values. When the Average Network Links Availability $ > A_{th} $, no protection is required; therefore, from here, we can interpret that there is no restorability ratio. Now, as the Average Network Links Availability $ \leq A_{th} $ backup D-cycles are required, requiring a higher number of protection paths. Since restorability is the ratio of the total number of protected paths to the total number of paths requiring protection (sum of the protected and unprotected paths), this ratio is only for the paths whose Availability is less than the threshold Availability.  Now, as the requirement for protection increases, the number of the protected paths also increases. Hence, the restorability ratio also increases. The redundant paths are provisioned to the working path till the $A_{pp}^{max} \geq A_{th} $. As the difference between Average Network Links Availability and $A_{th}$ progresses, the restorability ratio decreases due to the unavailability of the spectrum slots. Therefore, the denominator term (protected paths $+$ unprotected paths) starts dominating. This means there is an increase in the number of unprotected paths. After a certain $A_{th}$, it reaches a steady-state condition. As the load values increases, the steady-state condition is achieved earlier. 

The increase in the protection of number of lightpath requests can be done by increasing the capacity of each link in the network, multiple core-based networks or multiple band optical transmission networks. 

\section{Conclusion} 
The protection of an optical network is a crucial aspect for RSA of lightpath requests. In this paper, we formulate heuristics to assign protection to the new working path without disturbing existing traffic on all other routes. However, assigning protection to all the requests is impossible. Therefore, we use the Availability of the link and path as a parameter for protection. This paper uses two protection techniques, Dynamic Shared Backup Path and Spectrum Slots (DSBPSS), for the path protection and design Dynamic Cycles (D-cycles) for the link protection. Both the protection techniques use adaptive parameters, i.e., consecutive spectrum slots. The consecutive spectrum slots (i.e., Type-II algorithm) keep changing with the network conditions. We performed a detailed analysis of the proposed algorithms for NSFNET network topology. Based on the observation from the plots, the performance of DSBPSS and D-cycles are comparable. However, the switching speed of D-cycles is better than DSBPSS. Therefore, D-cycles can be considered for protection provisioning considering spectrum slots converter at the backup nodes. Whereas, if the network is not equipped with spectrum converters, then DSBPSS is an alternative solution for Routing, Spectrum Assignment and Protection Provisioning.

\end{document}